\documentclass[journal]{IEEEtran}

\usepackage{amsmath,amssymb,amsfonts}
\usepackage{algorithmic}
\usepackage{graphicx}
\usepackage{textcomp}
\usepackage{xcolor}
\usepackage{adjustbox}
\usepackage{placeins}
\usepackage{dblfloatfix} 
\usepackage{changepage} 
\usepackage{tikz,graphics,color,float,epsf,caption,subcaption}

\newtheorem{thm}{Theorem}
\newtheorem{lemma}[thm]{Lemma}
\newtheorem{cor}[thm]{Corollary} 

\usetikzlibrary{positioning}
\usetikzlibrary{shapes.geometric, arrows, positioning}

\usetikzlibrary{arrows.meta, calc, decorations.pathreplacing,
                ext.paths.ortho, positioning, quotes}
\makeatletter
\tikzset{phantom node/.code=\tikz@addoption{\expandafter\let\csname pgf@sh@boxes@\tikz@shape\endcsname\pgfutil@empty}}
\makeatother
\tikzset{
  shadowed node xshift/.initial=1.5ex, shadowed node yshift/.initial=1ex, shadowed node list/.initial={2, 1},
  pics/shadowed node/.default=\pgfkeysvalueof{/tikz/shadowed node list},
  shadowed node/.pic={
    \foreach[expand list] \elem in {#1}
      \scoped[transparency group, shadowed node calculation={\elem}]
        \node[style/.expand once=\tikzpictextoptions, phantom node,
              xshift={\elem*\pgfkeysvalueof{/tikz/shadowed node xshift}},
              yshift={\elem*\pgfkeysvalueof{/tikz/shadowed node yshift}}] (-\elem) {\tikzpictext};
    \node[alias=-0, style/.expand once=\tikzpictextoptions] () {\tikzpictext};},
  set shadowed node calculation parameter/.style={shadowed node calculation/.style={opacity={(#1-##1+1)/(#1+1)}}},
  set shadowed node calculation parameter=2,
  overshoot line to/.style={to path={($(\tikztostart)!-(#1)!(\tikztotarget)$)--($(\tikztotarget)!-(#1)!(\tikztostart)$)\tikztonodes}},
  edges have transparency group/.style={execute at begin to={\scope[transparency group,#1]}, execute at end to=\endscope}}

\begin{document}


\title{Ksurf-Drone: Attention Kalman Filter for Contextual Bandit Optimization in Cloud Resource Allocation}

\author{\IEEEauthorblockN{Michael Dang'ana, Yuqiu Zhang, Hans-Arno Jacobsen}

\IEEEauthorblockA{\textit{Electrical \& Computer Engineering} \\
\textit{University of Toronto}\\
Toronto, Canada \\
michael.dangana@mail.utoronto.ca, yuqiu.zhang@mail.utoronto.ca, jacobsen@eecg.toronto.edu
 }






}

\maketitle




\begin{abstract}
Resource orchestration and configuration parameter search are key concerns for container-based infrastructure in cloud data centers. Large configuration search space and cloud uncertainties are often mitigated using contextual bandit techniques for resource orchestration including the state-of-the-art Drone orchestrator. Complexity in the cloud provider environment due to varying numbers of virtual machines introduces variability in workloads and resource metrics, making orchestration decisions less accurate due to increased nonlinearity and noise. Ksurf, a state-of-the-art variance-minimizing estimator method ideal for highly variable cloud data, enables optimal  resource estimation under conditions of high cloud variability. \\

This work evaluates the performance of Ksurf on estimation-based resource orchestration tasks involving highly variable workloads when employed as a contextual multi-armed bandit objective function model for cloud scenarios using Drone. Ksurf enables significantly lower latency variance of $41\%$ at p95 and $47\%$ at p99, demonstrates a 4\% reduction in CPU usage and 7 MB reduction in master node memory usage on Kubernetes, resulting in a 7\% cost savings in average worker pod count on $VarBench$ Kubernetes benchmark.
\end{abstract}




\maketitle

\section{Introduction}
Containers have become the prevalent technology for deployment of applications and computation jobs in the cloud at scale due to their support for fine-grained controllability, fluid orchestration and scalability when compared to virtual machines \cite{zhang2023lifting} \cite{eismann2020microservices} \cite{luo2021characterizingAlibaba}. Wide adoption of containers has introduced \textit{cloud} and \textit{workload uncertainties} leading to increased risks of under- and over-provisioning cloud infrastructure \cite{kabir2021uncertainty} \cite{dean2013tail} \cite{luo2022power}. Unstructured and time-variant resource performance and cost relationships are prevalent in public cloud systems exacerbating uncertainties in cloud resource usage and inter-container traffic patterns \cite{awsSpotPricing} \cite{awsBurstIntances} \cite{shi2015clash}. \\

Temporal changes in the cloud environment including varying numbers of virtual machines, resource availability fluctuations and growth in user numbers introduce \textit{cloud}, \textit{workload} and \textit{resource variability} increasing uncertainty in orchestration decisions \cite{ProScale2023} \cite{autoscale2014}. Common workload variability incidents include flash crowds that refer to significant numbers of users rapidly accumulated in localized areas for short durations called flash events, or when popular sites are connected to smaller sites with substantial traffic spikes called flash-dot effects, causing reduced, lost and delayed packets due to congestion at the network layer \cite{DDoSFlashCrowd2011}. Flash crowds also provide a vector for security violations called Flash crowd attacks, where attack sources mask themselves in the flash crowd to execute direct denial of service attacks often using botnets to increase the severity of workload variability, making it difficult to minimize costs, optimize resource usage, and increasing the risk of Service Level Agreement (SLA) violations \cite{FlashCrowdEnsembleDetection2023althobaiti} \cite{yu2020microscaler}.  \\ 

Drone, an online resource orchestration framework is state-of-the-art for \textit{cloud} and \textit{resource uncertainty} mitigation for containerized clouds using Contextual Bandit techniques \cite{zhang2023lifting}. Drone requires minimal configuration for autonomous orchestration, avoiding the need for costly workload profiling by employing Guassian process-based contextual bandits to model uncertainty, demonstrating sublinear growth in convergence time for unlimited resource conditions of both public and resource-constrained private clouds \cite{ctxGBBandit}. Gaussian process models are susceptible to nonlinearity in data which can be effectively mitigated using Extended Kalman Filter-based techniques such as Ksurf, the state-of-the-art in resource estimation under conditions of high variability \cite{kfapplication} \cite{kfEKFSpaceZhang} \cite{KsurfConference}. \\

This work incorporates Ksurf and Drone for uncertainty mitigation under high \textit{workload} and \textit{resource variability} conditions evaluated using \textit{VarBench} Kubernetes benchmark \cite{KsurfConference}. We compare Ksurf and Gaussian process modeling for variability mitigation in orchestration decision-making scenarios using Drone and perform evaluations using non-recurring jobs for which bandit-based approaches are sub-optimal \cite{alipourfard2017cherrypick} \cite{ProScale2023}. \\

This work advances Drone by making the following contributions:
\begin{itemize}
    \item Proposes a novel Ksurf-Drone algorithm for cloud resource orchestration under conditions of environment uncertainty and \textit{variability}
    \item Proposes a novel KsurfNet algorithm for unsupervised tuning of Kalman filter parameters
    \item Provides a bounded analytical cummulative regret guarantee with exponential convergence for well-behaved systems
    \item Proposes a selection heuristic for KsurfNet context estimation and performs evaluation using the \textit{VarBench} Kubernetes cluster benchmark
    \item Compares Ksurf-Drone and Drone on container orchestration tasks under high \textit{cloud}, \textit{resource} and \textit{workload variability} using Compute Canada and public Google Cloud environments. \\
\end{itemize}

\section{Background and Related Work}
Contextual bandit algorithms are a new category of cloud resource orchestration techniques that avoid dependency on model pre-training, supervision and configuration inherent in the main cloud resource orchestration categories of heuristics, model analytics, and predictive methods \cite{zhang2023lifting} \cite{ctxGBBandit}. The contextual bandit is a variant of the Multi-Armed Bandit (MAB) problem that uses contextual information about environment variables in the form of uncertainties in the cloud, and is a discrete form of Bayesian Optimization (BO) that seeks to progressively build up an objective function model of an unknown system \cite{banditsintroduction}. \\

Container cloud orchestration is impacted by the prevalence of high workload variability and the need to ensure SLA and minimize costs in cloud computing, and techniques employed to address these issues often combine Bayesian optimization and multi-objective decision-making \cite{BanditsNoiseConfSetsLattimore} \cite{rambo2021}. State-of-the-art bandit algorithms for system orchestration include Cherrypick, which lacks convergence guarantees for its acquisition function, and Accordia, both of which study the virtual machine configuration selection problem under recurring analytical jobs with predictable workloads \cite{alipourfard2017cherrypick} \cite{liu2019accordia}. The container configuration selection problem is qualitatively complex requiring fine-grained continuous control well-suited to Drone, which uniquely generalizes to workload variations and variable cloud settings due to lack of pre-training, a key step for Cherrypick and Accordia, and a feature shared with RAMBO that uses multi-objective BO to solve the performance-cost trade-off between SLA and Total Cost of Ownership (TCO) for the customer in multi-microservice scenarios, providing pareto-optimal performance guarantees \cite{rambo2021}. \\

Trade-offs inherent to multiple objective optimization, including minimization of CPU usage, memory and latency, elicit the use of multi-objective multi-armed bandits where the effects of decisions are modeled as vectors of objectives over which no assumptions can be made about optimal utility functions \cite{MoRLMABGP}. Whereas the strength of Gaussian process algorithms is their ability to learn arbitrary utility functions, constraints are useful in providing contextual awareness about the environment within utility functions, enabling accurate action selection at each iteration, leading to the development of contextual bandit algorithms \cite{ctxGBBandit}.\\

Gaussian process regression (GP) is widely used in contextual bandit algorithms for non-parametric Bayesian modeling of unknown systems \cite{ctxGBBandit}. In Drone, GP is used to select actions for exploration, with updates after action exploitation in private and public cloud scenarios \cite{zhang2023lifting}. GP regression is a batch processing technique with appreciable polynomial execution cost of $O(n^3)$, in contrast to Kalman filter methods that are online, low cost, ideal for nonlinear data where system model and transition functions are well defined with sub-polynomial cost \cite{KFGP} \cite{Ksurf+}. Ksurf (AKF-PCA) provides further benefits over GP for Drone by employing noise mitigation using dimensionality reduction and modeling of temporal relationships in data through the attention mechanism adopted from state-of-the-art Transformer neural network architecture; where GP struggles with abrupt transitions and non-stationarities, data-driven Kalman filters can adapt quickly to these regime shifts \cite{GP2014automatic} \cite{KsurfConference} \cite{KalmanNet2022}. \\


GP-based approaches enable high approximation accuracy and analytic uncertainty characterization in contextual decision-making algorithms such as neural network-accompanied Gaussian processes (NN-AGP) \cite{nnAGPB}. When decision and reward functions are derived from continuous sets, estimating the reward function while taking uncertainty and accuracy into account is a significant challenge often tackled using GP, which remains susceptible to time-varying reward variables and graph-structured contextual variables \cite{opolka2022adaptive}. Similar to neural network-based model completion schemes like KalmanNet, NN-AGP performs model completion by approximating the unknown reward function and allowing GP to compute the decision variable, leading to state-of-the-art approximation accuracy and explicit GP uncertainty quantification \cite{nnAGPB} \cite{KalmanNet2022}. \\

\begin{figure*}[!htpb]
  \centering
  \includegraphics[width=0.79\textwidth]{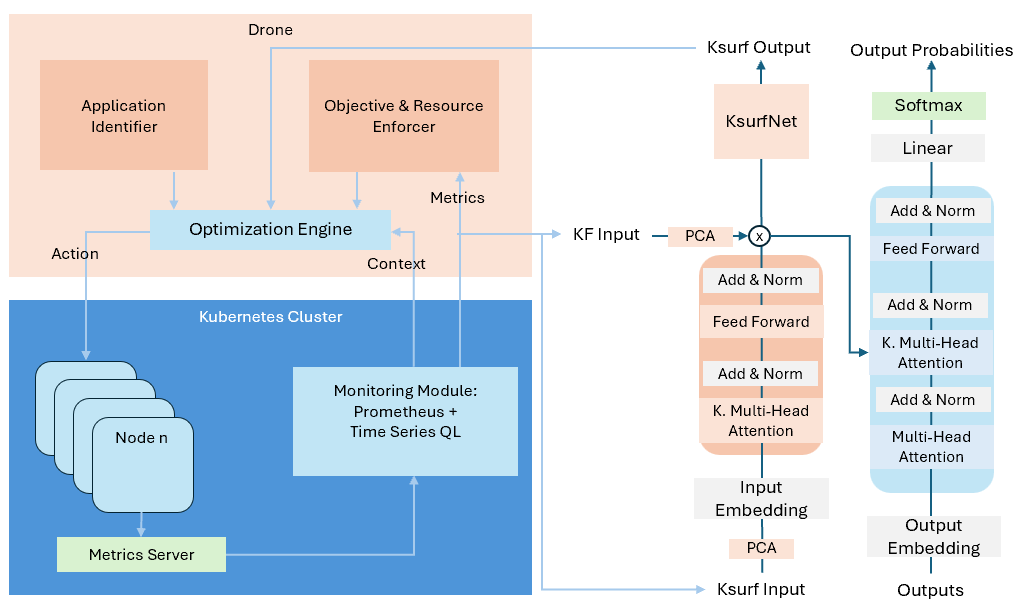}
  \caption{Ksurf-Drone Optimization Architecture}
  \label{fig:ksurf_drone}
\end{figure*}

Ksurf improves upon these approaches by incorporating Kalman filter-based online learning which provides robustness under high variability while avoiding manual configuration selection and the complexity cost of offline batch methods, well adapted to time-varying variable estimation \cite{KsurfConference}. Ksurf-Drone combines these approaches into a dynamic contextual bandit optimization module, outperforming GP-based optimization regarding resource variability with convergence guarantees, and offers more scalability for resource-constrained scenarios due to lower complexity. Kalman filters have demonstrated the capability for inference of graph-structured models offering the potential to tackle a key limitation of GP \cite{GSPKalmanNet}.  \\

\section{System Architecture}
\begin{figure}[h]
\centering
\includegraphics[clip,width=1.09\linewidth]{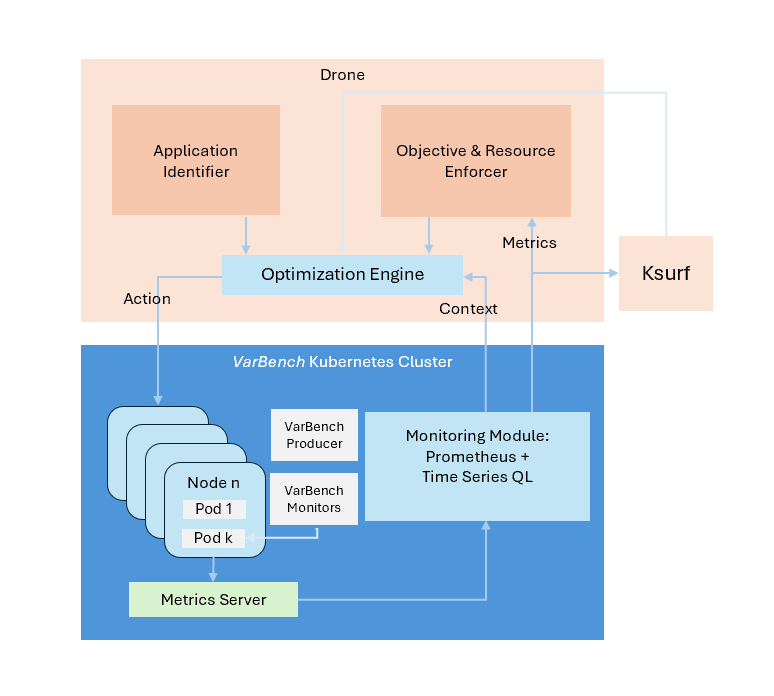}
\caption{Ksurf and Drone in \textit{VarBench} Kubernetes Cluster}
\label{fig:drone_kubernetes}
\end{figure}

This work incorporates Ksurf as an optimization function for the Drone contextual bandit for robust noise reduction and modeling of temporal relationships to improve accuracy \cite{KsurfConference}. The Drone optimization module supports configurable optimization functions and function selection as shown in Figure~\ref{fig:drone_kubernetes}. \\ 

\begin{figure}[H]
  \centering
  \includegraphics[width=0.47\textwidth]{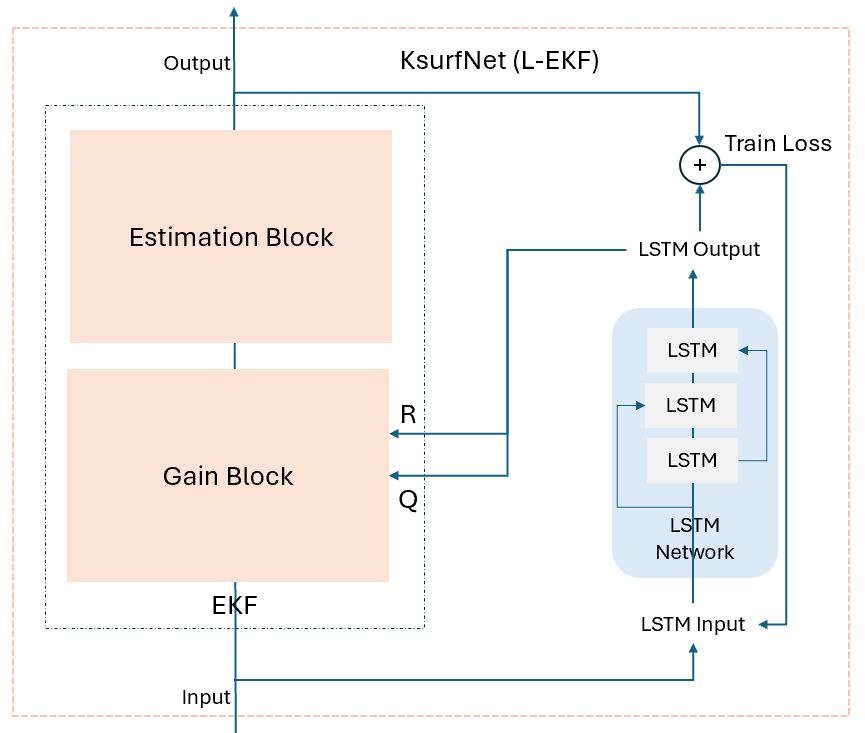}
  \caption{KsurfNet Component Architecture}
  \label{fig:ksurf_lstm_ekf}
\end{figure}

Drone consists of a Monitoring Module responsible for collecting performance metric including CPU, RAM, network bandwidth, and cloud context using Prometheus metrics system \cite{prometheus}. The Application Identifier determines the application type, which can be user-specified, for targeted orchestration decisions \cite{zhang2023lifting}. The Objective and Resource Enforcer sets the objective for the optimization engine, where users can specify model parameters, performance-cost coefficients and resource limits. The Optimization Engine executes the optimization process, receiving and processing performance and contextual metrics and exploring resource orchestration actions iteratively, which are exploited through the Kubernetes API \cite{k8sResources}. Ksurf is executed as the objective function model by the Optimization Engine in private and public cloud algorithms during the exploration action selection and exploitation posterior-update steps in Figure \ref{fig:ksurf_drone}. \\

Ksurf includes an attention network which follows the standard design with $L=4$ layers, each with $N$ tokens, dimension $d=256$, feed-forward $d_{ff}=4$, attention heads $h=4$, attention dimension $d_{k}=d_{v}=256$ and dropout rate of 0.25. The KsurfNet Long Short-Term Memory (LSTM) subnetwork has $L_s=10$ sequence length, $N=16$ batch size, $\lambda=0.001$ learning rate, with an Adam optimizer, $L=2$ hidden layers of size $N_h=64$ tokens, fully connected output layer with $R$ and $Q$ outputs trained over $\epsilon \in \{1,30\}$ epochs. \\

\subsubsection{Ksurf EKF Parameterization using KsurfNet}
A key contribution of this work is KsurfNet, that uses a LSTM network to learn initial noise parameters for the EKF estimator (L-EKF) using a training subset of the inference data set, a data-driven model-based approach that has proven effective for state estimation with only partial knowledge of the underlying model parameters as shown in Figure \ref{fig:ksurf_lstm_ekf} \cite{KalmanNet2022}. The observation and process R and Q matrices are estimated using an offline approach where EKF input data feeds into the LSTM and the EKF estimate error is used as the LSTM loss function in Figure \ref{fig:ksurf_lstm_ekf}. A Grid Search approach is used to tune the learning rate $\lambda = 0.001$ and the number of epoch $\epsilon = 10$ hyperparameters \cite{LstmArimaHyperParameterGrid}. \\

\subsubsection{Ksurf Kalman Filter Algorithm}
The observation measurements and Gaussian white noise are expressed as vectors $z_{k}$ and $u_{k}$. The process state and Gaussian white noise vectors express the resource state represented by $x_{k}$ and $w_{k}$ \cite{KFComplexity}. The initial process state $x_{0}$ vector is measured from the resource metrics. Define $H$ as the observation matrix which relates the workload and resource state, and $A$ as the state transition matrix which updates the resource state. The observation and state update equations are

\begin{equation}
 \begin{aligned}
    & z_{k} = Hx_{k} + u_{k-1} \\
    & x_{k} = Ax_{k-1} + w_{k}
  \end{aligned}
\end{equation}

The covariance matrix $P$ captures information about the measured error in the state vector initialized as $P^{-}_{k} = E[e^{-}_{k}.e^{-T}_{k}]$ and $P_{k} = E[e_{k}.e^{T}_{k}]$ where $E[.]$ is the expectation operator. \\

The Kalman gain matrix $K_{k}$ is used with measurement estimation error to compute the resource state update:

\begin{equation}
\begin{aligned}
    & R_{k} = E[u_{k}u^{T}_{k}]\\
    & K_{k} = P^{-}_{k}H^{T}(HP^{-}_{k}H^{T} + R_k)^{-1} .\\
\end{aligned}
\end{equation}

\subsubsection{Ksurf Contextual Bandit Algorithm}
Contextual Bandit Problems involve the reward $r_t(a_t)\in [0,1]$ of action $a_t$ on round $t$, where the agent observes $K$ feature vectors $c_t \in \mathbb{R}^d$ for each action with $\|c_t\|\leq 1$ and selects $a_t$ to receive $r_t(a_t)$ \cite{BanditsCtxLinearChu11a}. It is assumed that there exists $\theta^* \in \mathbb{R}^d$ with $\|\theta^*\|\leq1$ such that $\mathbf{E}[r_t|c_t]=c_t^T\theta^*$, the goal is to minimize \textit{regret} defined as

\begin{equation}
\begin{aligned}
    & R_t = \sum_{t=1}^Tr_t(a_t^*)-\sum_{t=1}^Tr_t(a_t)
\end{aligned}
\end{equation}

The contextual bandit agent chooses a scaling action $a_t$ and receives a reward such as inverse latency and throughput $r_t(a_t) = g(a_t, c_t) + s_t$ such that reward noise $s_t$ is a Gaussian with standard deviation $\sigma_t$  \cite{banditsintroduction}. The Drone contextual bandit takes Gaussian Process (GP) state as input $c_t=\psi(z_t)$ with GP noise variance $\sigma_t^2$, while Ksurf provides a filtered state variable enabling contextual bandit $c_t^{KF}=\psi(x_t)$ with noise reduction and temporal attention filtering \cite{BanditsNoiseConfSetsLattimore} \cite{KsurfConference}. \\

The Ksurf context feature error is $\epsilon_t=\psi(\hat{x}_t)-\psi(x_t) \approx J_t(\hat{x}_t-x_t)=J_te_t$ effective contextual bandit input noise is $s^{KF}_t=J_te_t+s_t$ with variance $(\sigma_t^{KF})^2=tr(J_tP_tJ_t^T)+\nu^2_t$ in contrast to the GP-based $\sigma_t^2 \approx tr(J_tH^{-1}_tR_t(H^{-1}_t)^TJ_t^T) + \nu^2_t$ where $z_t \approx H_tx_t$ \cite{BanditsEpsilonShrink2011abbasi}. Kalman filter error convariance $P_t$ converges at steady state \cite{kfapplication} 

\begin{equation}
\begin{aligned}
    & P_t\ll H_t^{-1}R_t(H_t^{-1})^T \\
    & \implies tr(J_tP_tJ_t^T) \ll tr(J_tH^{-1}_tR_t(H^{-1}_t)^TJ_t^T) \\
    & \implies (\sigma_t^{KF})^2 \ll \sigma_t^2 \\
\end{aligned}
\end{equation}

The regret improvement by variance reduction factor $\rho_t=\frac{(\sigma_t^{KF})^2}{\sigma_t^2} \ll 1$ leads to improved contextual bandit exploration efficiency and exploitation stability enabling better resource orchestration performance \cite{BanditRegretVariance2024jia2024} \cite{BanditsNoiseConfSetsLattimore}. Since $r_t(a_t)$ is Lipshitz continuous we get the following bounded regret guaranteeing exponential convergence for well-behaved systems \cite{krener2002convergence}: \\
\begin{lemma}\label{Regret lemma}
\begin{equation}
\begin{aligned}
    & |R_t| \leq L\|c^*_t-c_t\| \leq L\|\psi(\hat{x}_t) - \psi(x_t)\| \\
    & \implies |R_t^{KF}| \leq L\|J_tP_tJ_t^T+\gamma^2_t\| \\
    & \implies |R_t^{KF}| \leq \sqrt{\rho}L\|J_tH^{-1}_tR_t(H^{-1}_t)^TJ_t^T + \gamma^2_t\| \approx \sqrt{\rho}|\epsilon_t| \\
    & \implies |R_t^{KF}| \leq \sqrt{\rho}|R_t| \\
\end{aligned}
\end{equation}
\end{lemma} 

Modelling latency as a Lipshitz continuous function of the action and context, it follows that $l_t=f(a_t,c_t)+\chi_t \implies |l^{KF}_t| \leq \sqrt{\rho}|l_t|$. A corollary of this is that since percentile latency is proportional to the standard deviation, and the p99 latency $l_{99,t} = E[l_t] + 2.33\sigma_t$  \cite{QueuesLipshitzWhitt2018time} \\

\begin{cor}\label{Latency Corollary}
\begin{equation}
\begin{aligned}
    & l_{99,t}^{KF} = E[l_t] + 2.33\sigma_t^{KF} \\
    & \implies |l_{99,t}^{KF}| = |E[l_t] + 2.33\sqrt{\rho_t}\sigma_t| \leq \sqrt{\rho_t}|l_{99,t}| \\
\end{aligned}
\end{equation}
\end{cor}

\subsubsection{Complexity Analysis}
The complexity of GP is $O(n^3)$ while Ksurf has a complexity of $O(m^3 + LN^2d)$ where $m$ is the most recent measurement size, $L,N,d$ are attention network layer, token and query size, and $n$ the data set size is the largest parameter $n > N \gg m$. Due to the need to perform regression over the entire data set at each iteration, GP is inherently more computationally intensive than Ksurf, however, this is apparent on very large data sets and grows as $n$ increases \cite{gregor2015draw} \cite{wenger2024computation}. \\

\section{Evaluations}
The evaluation experiments include a Ksurf and Drone baseline comparison using the Google Charlie trace benchmark, Drone benchmark experiments, KsurfNet ablation study on Google Cloud, $\mu Bench$ benchmark experiments using Drone for resource allocation and Prometheus for metric retrieval, computation and persistence on $\mu Bench$ application, and $VarBench$ benchmarks on Compute Canada cloud \cite{googleBenchmark} \cite{computeCanada}. \\

The experiment setup includes 10 4-core x86 64bit Intel Platinum 6548Y+ 2.5GHz 60MB cache (Broadwell) CPU and 15.4GB DDR5 memory compute nodes running Ubuntu Linux, Kubernetes 1.32 and Apache Kafka 2.8.2 in Kubernetes pod containers on Arbutus Compute Canada cloud, and Intel Broadwell 4vCPU 16GB 2.25 GHz x86-64 e2-standard-4 Ubuntu 22.04.2 100GB Google Cloud Compute Instances for $\mu Bench$ experiments. Experiment datasets include 1-dimensional resource traces with temporal sliding windows used as a second dimension. The Attention neural networks are trained using a 10 thousand-value training set selected from the respective experiment time series historical data. Ksurf auto-scaler (KF), Drone (DR), Ksurf-Drone (DRKF), KsurfNet-Drone (DRRQ), threshold crossing auto-scaler (TH) and non-threshold crossing baseline (NA) perform orchestration in the experiments. \\

\subsubsection{Ksurf Attention Network Parameters and Training Costs}
The Ksurf attention network has higher runtime cost than the Extended Kalman filter  with approximately with 5x higher execution time on 10 thousand data points and 40\% higher memory usage than the EKF due to the CPU-based attention sub-network forward and backward pass latency and matrix computation footprint during training and inference. The costs are lower than on the GP regression model due to the attention layers being smaller than a full end-to-end GP covariance kernel matrix \cite{GPKalmanRTS}. \\

\begin{figure}[H]
\centering
\includegraphics[clip,width=1.0\linewidth]{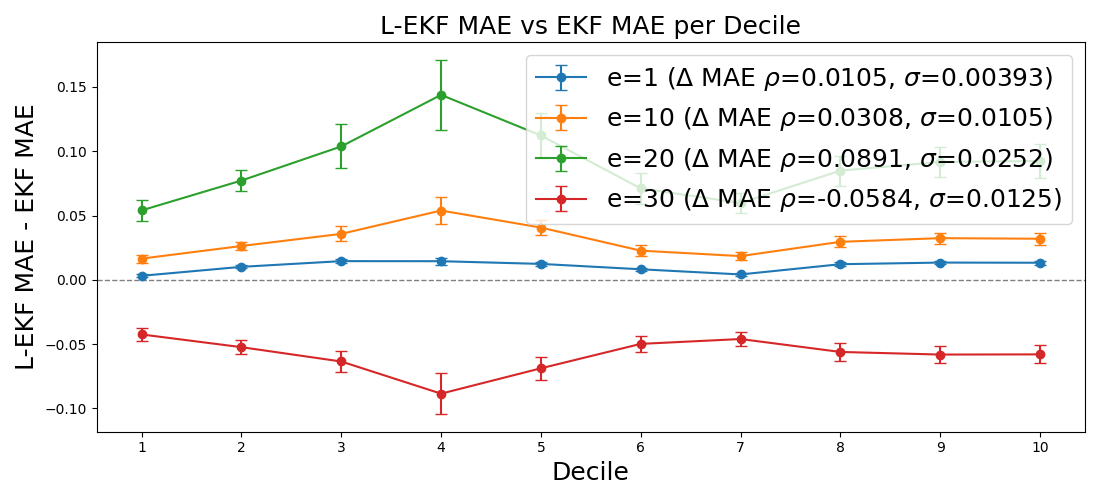}
\caption{KsurfNet MAE comparison to EKF MAE}
\label{fig:ksurf_epoch_delta_lines}
\end{figure}

\subsection{KsurfNet Baseline Comparison}
Grid search is used to determine the learning rate $\lambda=0.001$ and number of epochs $\epsilon = 30$ for optimal state estimation. The data set is chunked into deciles for state estimation using the KsurfNet and baseline models, the inference error difference between predicted error and baseline error $\Delta$ is computed (negative is better) with $95\%$ confidence intervals as shown in Figure \ref{fig:ksurf_epoch_delta_lines}. At $\epsilon=30$ KsurfNet shows the best performance with the lowest $\Delta$ and best non-zero confidence intervals, there is moderate $\Delta$ variance, however, increasing $\epsilon$ does not guarantee further gains as there is no definitive trend.  \\

\begin{figure}[H]
\centering
\includegraphics[clip,width=1.0\linewidth]{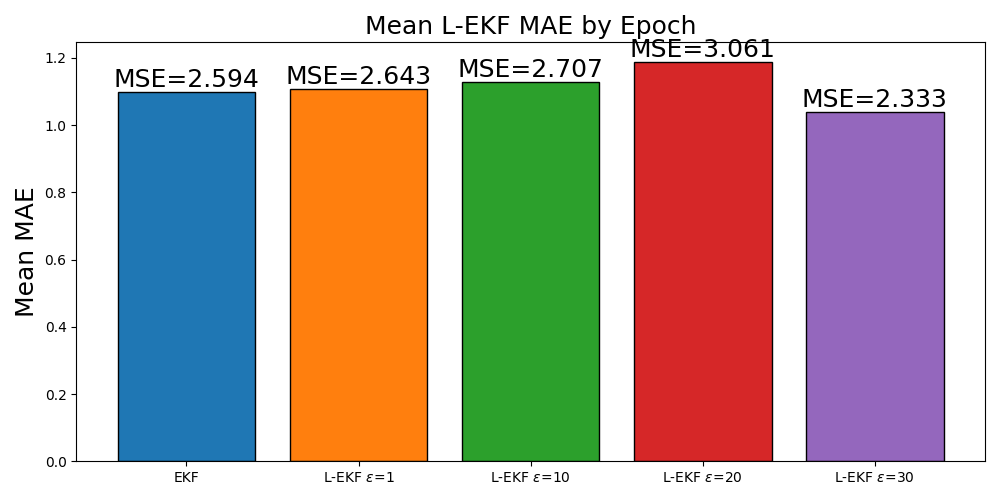}
\caption{KsurfNet MAE by Epoch}
\label{fig:ksurf_epoch_bar_mae}
\end{figure}

Training Mean Absolute Error (MAE) between EKF and KsurfNet L-EKF $\epsilon \in \{1, 10, 20, 30\}$ is compared where $\epsilon=30$ has the best performance as shown in Figure \ref{fig:ksurf_epoch_bar_mae}. Training L-EKF with low $\epsilon$ underfit noise parameters making L-EKF less accurate than the baseline, at $\epsilon = 20$ learned Q and R appear to be overfit, while at $\epsilon=30$ L-EKF shows error improvement over the baseline at all intervals during inference, matching the training performance shown in Figure \ref{fig:ksurf_epoch_delta_lines}. This enables KsurfNet to improve Ksurf by reducing prediction error through more accurate Q and R Gaussian noise parameters. Note that due to the training accuracy being non-correlated to $\epsilon$, the error plot shows smaller $\epsilon$ or limited fine-tuning data leads to underfit models common to data-driven model-based methods \cite{MAMLKalmannet2025}. \\

\begin{figure}[H]
\centering
\includegraphics[clip,width=1.0\linewidth]{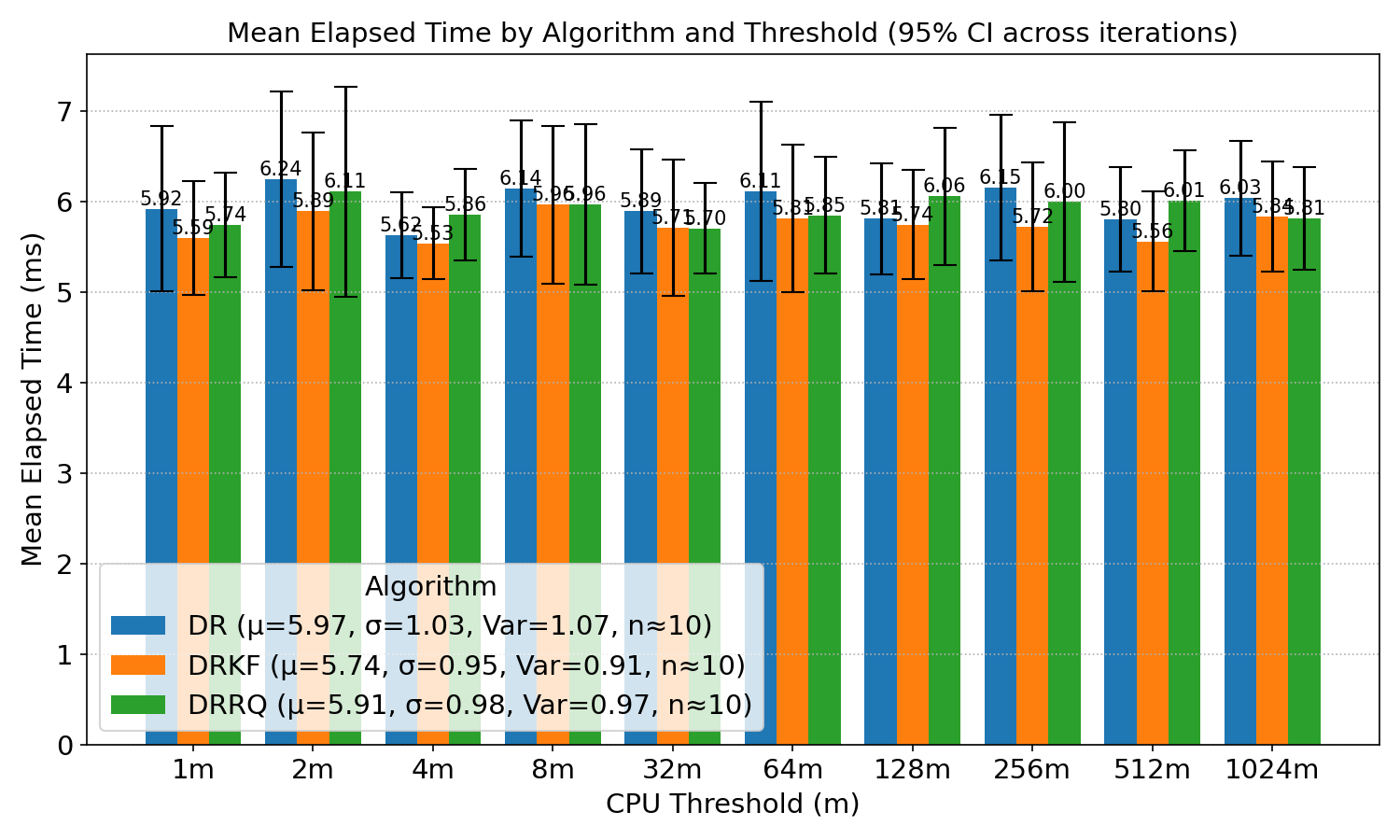}
\caption{Google Cloud Mean Request Latency by Threshold}
\label{fig:gcloud_mean_elapsed_time}
\end{figure}

\subsection{Google Cloud Benchmark Experiments}
Google cloud benchmark experiments involve three contextual bandit algorithms including Drone with Gaussian Process Regression (DR), Drone with Ksurf (DRKF) and Drone with KsurfNet (DRRQ) with a $\mu Bench$ serial 10-service application serving a stochastic workload on a custom Kubernetes cluster running on 8 Google Cloud instances, analyzing application request latency with varying CPU threshold where each experiment is ran 10 times per algorithm to establish confidence intervals \cite{muBench}. The Kubernetes cluster and $\mu Bench$ benchmark use the same node manifests as the Compute Canada private cloud experiments shown in Figure \ref{fig:varbench_cluster}. \\

\begin{figure}[H]
\centering
\includegraphics[clip,width=1.0\linewidth]{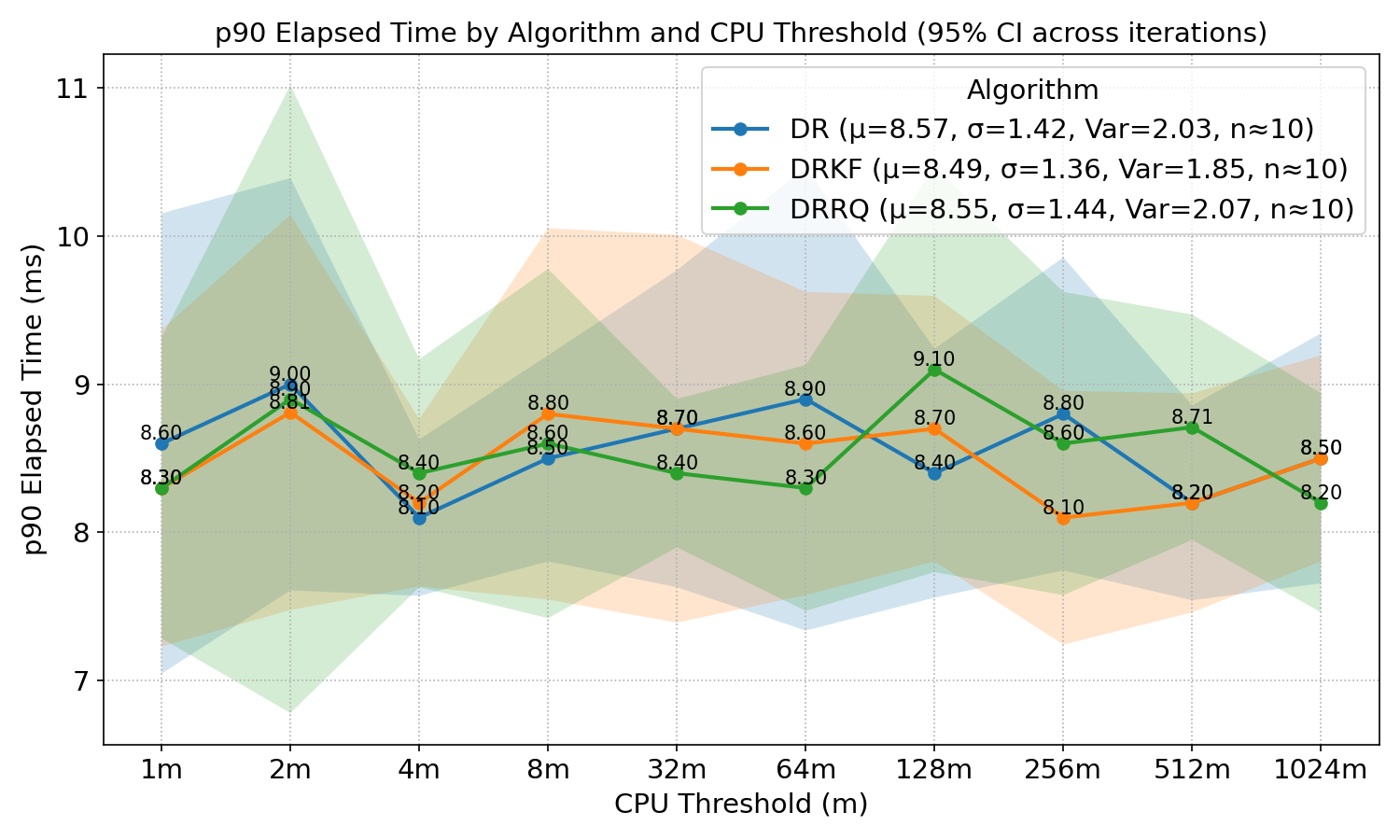}
\caption{Google Cloud p90 Request Latency by Threshold}
\label{fig:gcloud_p90_mean_elapsed_time}
\end{figure}

P90, p95, p99 and average latency are measured throughout the workload request duration reported as the mean, standard deviation and $95\%$ confidence interval. DRKF has a $4\%$ lower mean latency, $8\%$ lower latency standard deviation and wider confidence intervals than the next best DR algorithm over all thresholds demonstrating the effectiveness of Ksurf variance reduction on DRKF resource orchestration performance as shown in Figure \ref{fig:gcloud_mean_elapsed_time}.  \\

\begin{figure}[H]
\centering
\includegraphics[clip,width=1.0\linewidth]{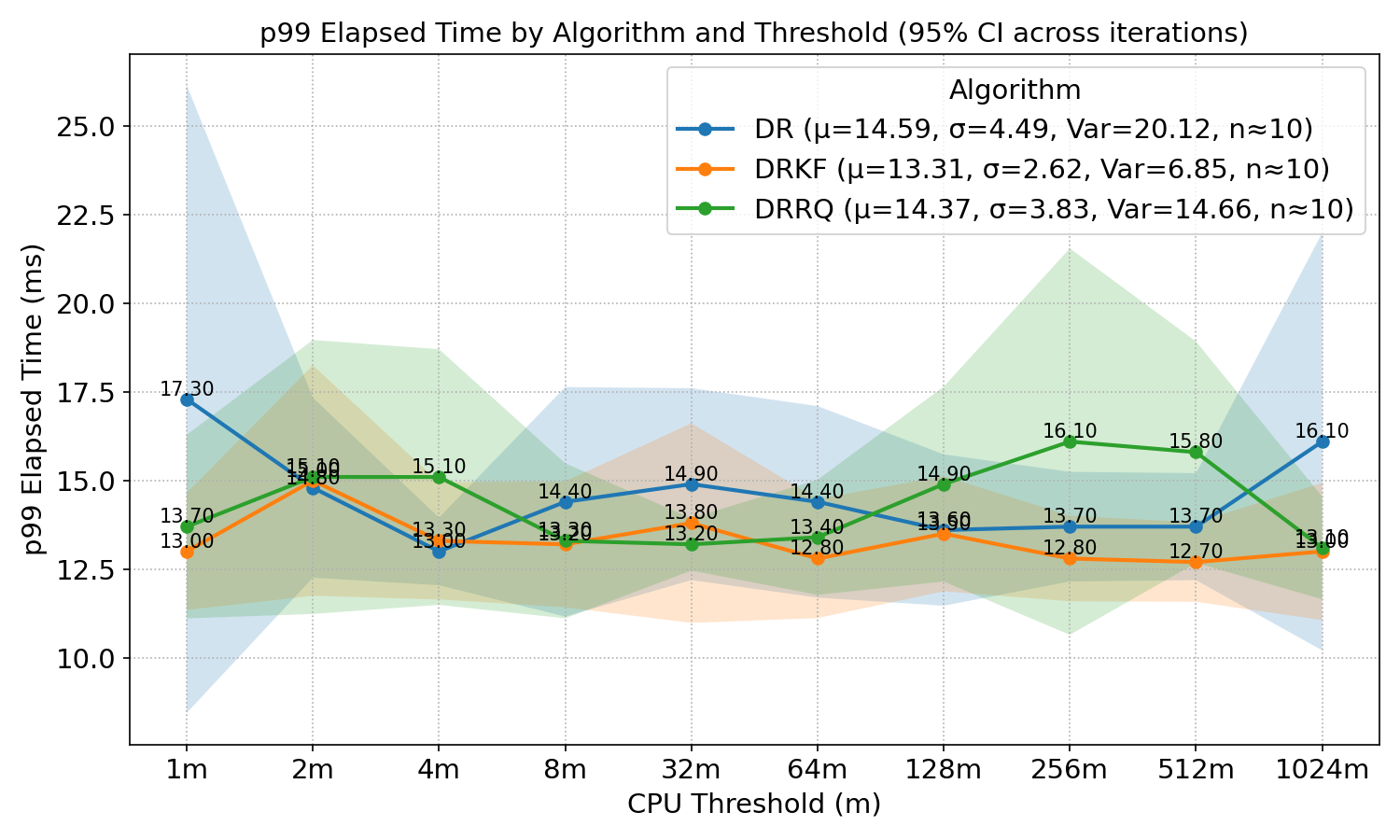}
\caption{Google Cloud p99 Request Latency by Threshold}
\label{fig:gcloud_p99_mean_elapsed_time}
\end{figure}

\begin{figure}[H]
\centering
\includegraphics[clip,width=1.0\linewidth]{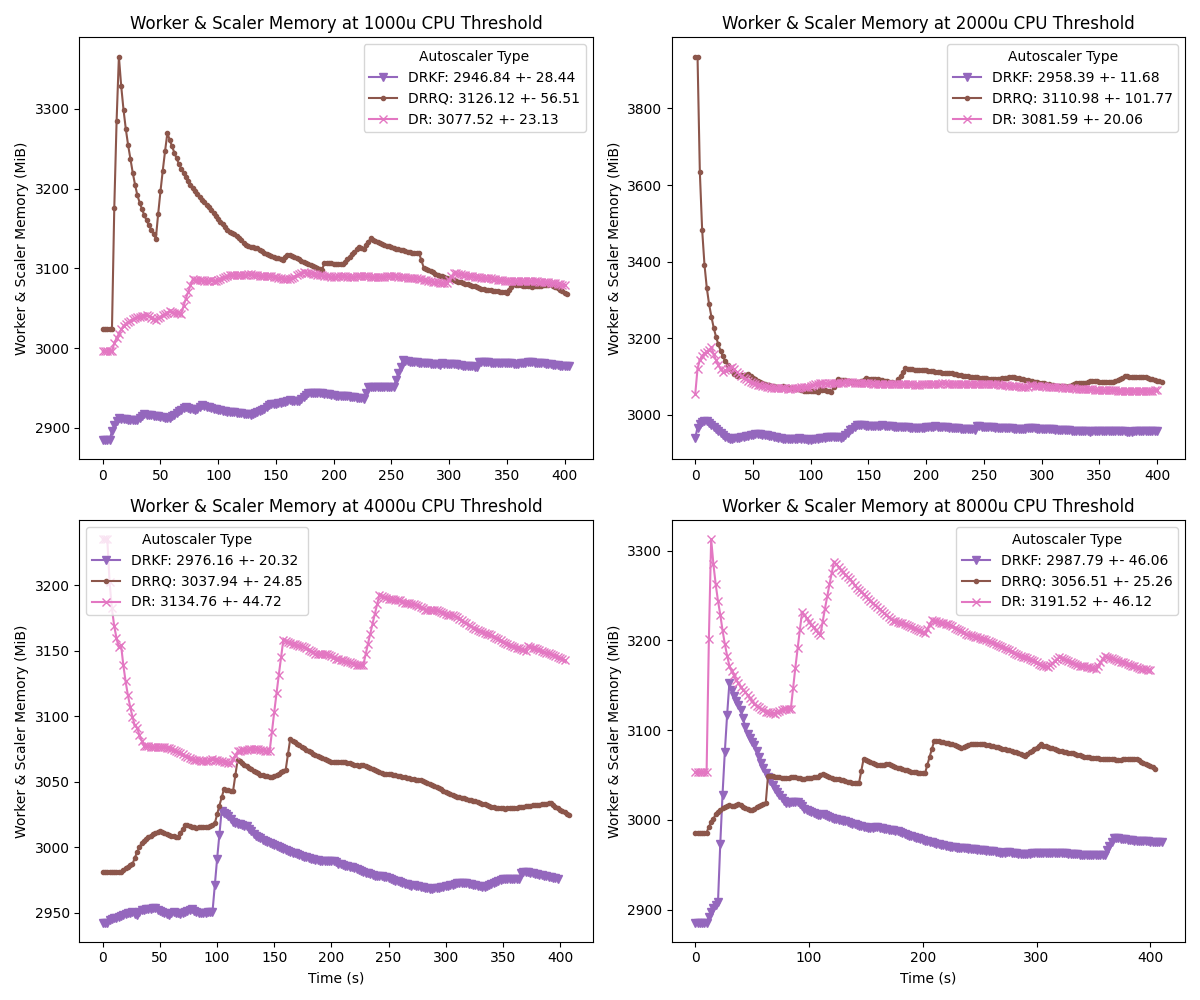}
\caption{Google Cloud Worker and Scaler Memory Usage by Threshold}
\label{fig:gcloud_ksurf_drone_node_metrics_summary}
\end{figure}

\begin{figure}[H]
\centering
\includegraphics[clip,width=1.0\linewidth]{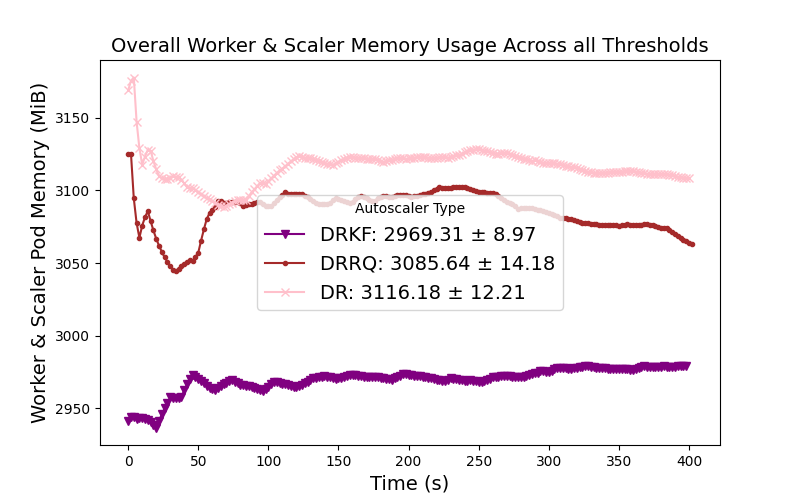}
\caption{Google Cloud Worker and Scaler Memory Usage Summary}
\label{fig:gcloud_ksurf_drone_worker_memory_all}
\end{figure}

Node memory usage is measured on the $VarBench$ Kubernetes cluster over the duration of the workload as shown in Figure \ref{fig:gcloud_ksurf_drone_worker_memory_all}. DRKF demonstrates a $5\%$ reduction in worker and scaler pod memory usage and $27\%$ reduction in memory usage standard deviation. This is due to the reduced memory usage of DRKF on the scaler node because of the memory efficiency of Ksurf in comparison to GP, and is consistent in all experiment thresholds as demonstrated in Figure \ref{fig:gcloud_ksurf_drone_node_metrics_summary}.  \\

\begin{figure}[H]
\centering
\includegraphics[clip,width=1.0\linewidth]{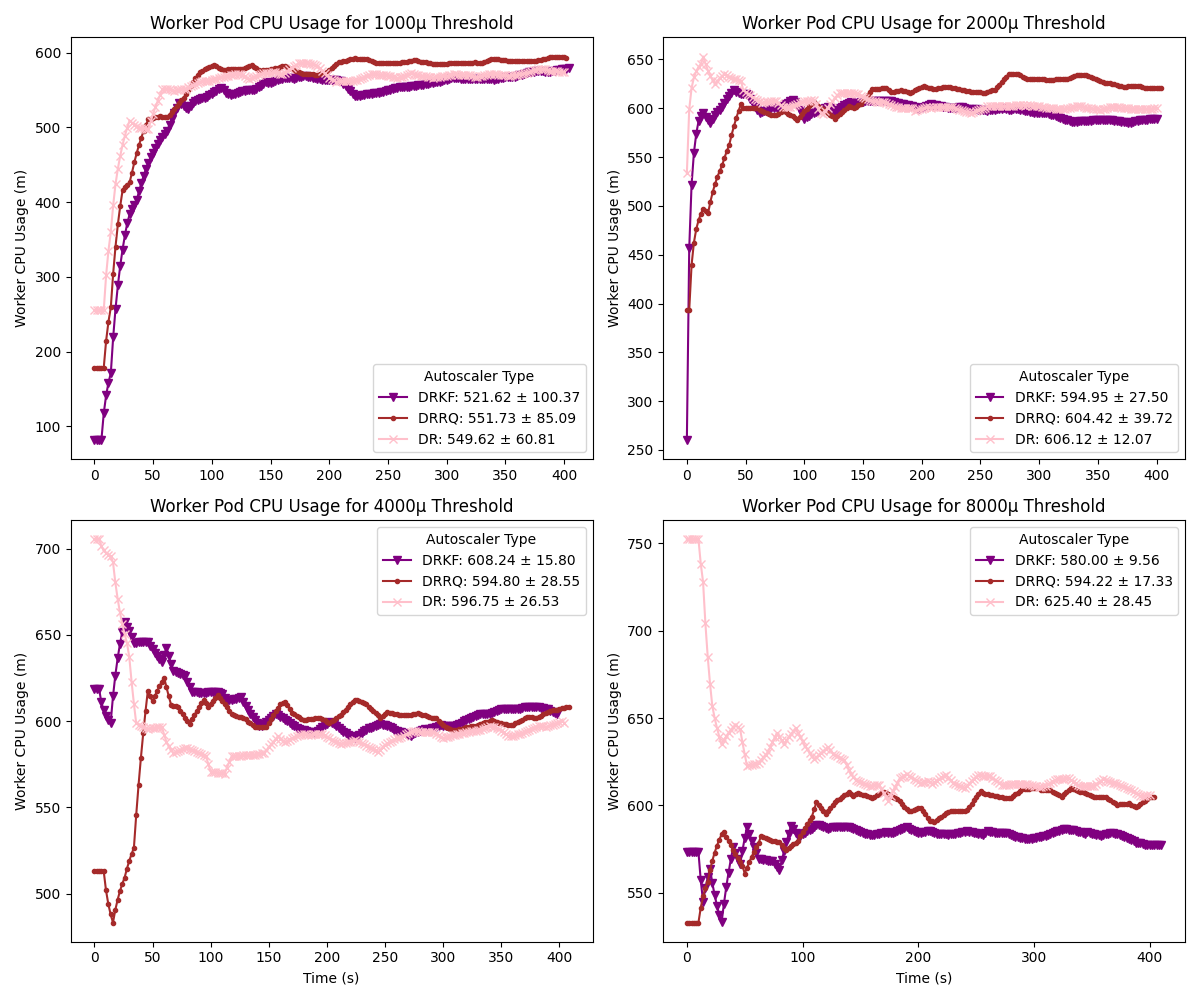}
\caption{Google Cloud Worker CPU Usage by Threshold}
\label{fig:gcloud_ksurf_drone_worker_cpu_512}
\end{figure}

\begin{figure}[H]
\centering
\includegraphics[clip,width=1.0\linewidth]{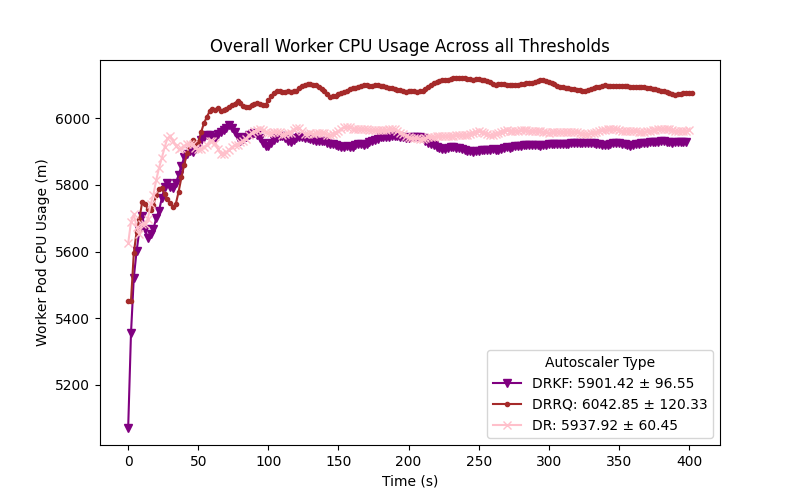}
\caption{Google Cloud Worker CPU Usage Summary}
\label{fig:gcloud_ksurf_drone_worker_cpu_all}
\end{figure}

In addition, pod CPU usage is measured during the workload duration as shown in Figure \ref{fig:gcloud_ksurf_drone_worker_cpu_all}. DRKF demonstrates a $1\%$ reduction in mean worker pod CPU usage and modest increase in standard deviation. This is caused by the higher worker CPU usage in high threshold experiments with the trend of high CPU usage at the beginning of the experiment which reduces as the experiment progresses due to Drone scaling, with higher peaks observed as thresholds increase as shown in Figure \ref{fig:gcloud_ksurf_drone_worker_cpu_512}. DRKF is better able to mitigate these high demand scenarios due to lower computation complexity in the on-line Ksurf algorithm enabling faster response to short-duration nonlinear spikes in measurements \cite{GPKalmanRTS}.  \\

A key observation from these experiments is the variability in node and pod metrics with time of day, and how this impacts the experiment methodology. Due to the significant variance in mean memory and CPU usage under the same workload, experiments on different auto-scalers have to be performed within short durations of each other in randomized order to minimize variability effects due to daily patterns in the Google Cloud Compute Engine environment \cite{GloudAwsVariabilty}. \\

\subsubsection{KsurfNet Ablation and Bandit Context Selection Heuristic}
The performance of data-driven KsurfNet is demonstrated by DRRQ which is on average intermediate between DR and DRKF with a $1\%$ lower mean request latency and $9\%$ lower latency variance in the Google Cloud $\mu Bench$ benchmark experiments, better than the DRKF in the baseline comparison in Figure \ref{fig:ksurf_epoch_delta_lines}. DRRQ learns both R and Q matrices from data while DRKF loads the pre-selected R and Q matrices from the configuration specification. DRRQ demonstrates $0.3\%$ and $2\%$ lower mean p90 and p99 latency, and has a $0.2\%$ slightly higher p95 mean latency than the DR algorithm as shown in Figure \ref{fig:gcloud_p90_mean_elapsed_time}, Figure \ref{fig:p95_elapsed_time} and Figure \ref{fig:p99_elapsed_time}. The tail benefits of DRRQ are demonstrated in the comparison of the mean and variance of latency p99 with p90 and p95 where the comparative performance increases significantly at p99 as the model latency $l_t=h(a_t,x_t)+\nu_t$ and $\rho_t$ is bound by $P_t$ which converges more quickly with well-tuned L-EKF noise parameters. \\

Although KsurfNet tuning demonstrates performance improvements in DRRQ in baseline comparison experiments as shown in Figure \ref{fig:ksurf_epoch_delta_lines}, it does not improve upon the performance of DRKF in Google cloud experiments due to the offline tuning phase that is susceptible to non-stationarity in the noise characteristics on Google cloud, a problem that warrants further examination in a future study due to time and resource constraints of this research effort. KsurfNet is best employed in an online setting where the LSTM training costs can be amortized by significant gain in the regret variance reduction factor $\rho_t\geq \rho_{min}$. For practical offline usage, the following heuristic guarantees DRRQ performance equivalent to DRKF or better for an empirically determined configurable parameter $\rho_{min}=0.1$.

\[
  c^{KF}_t =  \{
\begin{array}{ll}
    \psi(\hat{x}_{EKF}), & |\rho_{t-1}| \leq |\rho_{min}| \\
    \psi(\hat{x}_{L-EKF}), & |\rho_{t-1}| > |\rho_{min}|  \\
\end{array}
\}
\]

\begin{figure}[H]
\centering
\includegraphics[clip,width=1.0\linewidth]{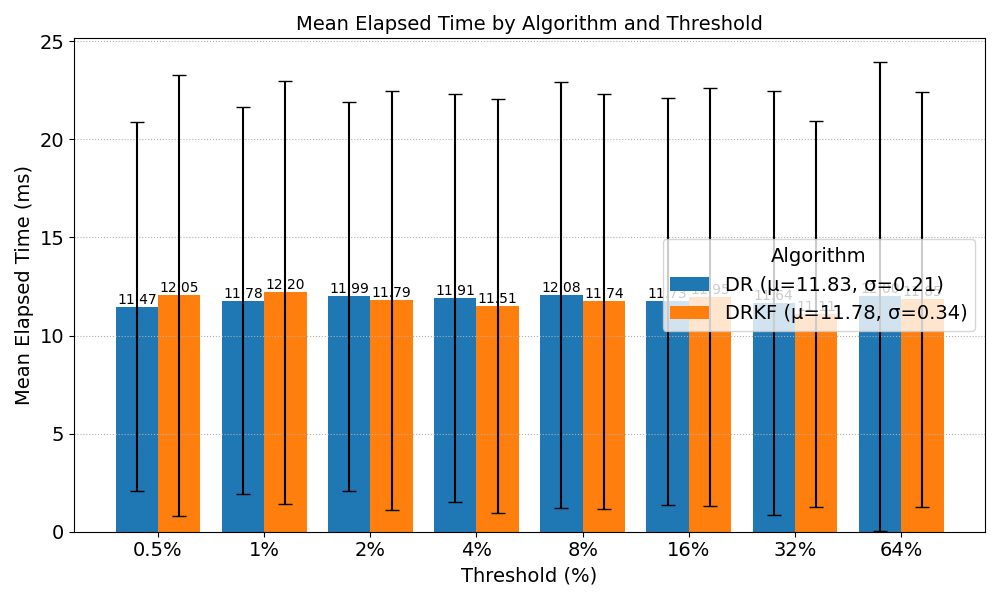}
\caption{Compute Canada $\mu$Bench Mean Request Latency By Threshold}
\label{fig:mean_elapsed_time}
\end{figure}

\subsection{$\mu$Bench Benchmark Experiments on Compute Canada}

$\mu Bench$ is a microservice benchmark for Kubernetes application performance modelling using stochastic, trace workloads \cite{muBench}, and simulates real-world scenarios using synthetic application workload models including DeathStarBench and Teastore \cite{muBenchDeathStar}. \\

\begin{figure*}[h]
\centering
\includegraphics[clip,width=1.0\linewidth]{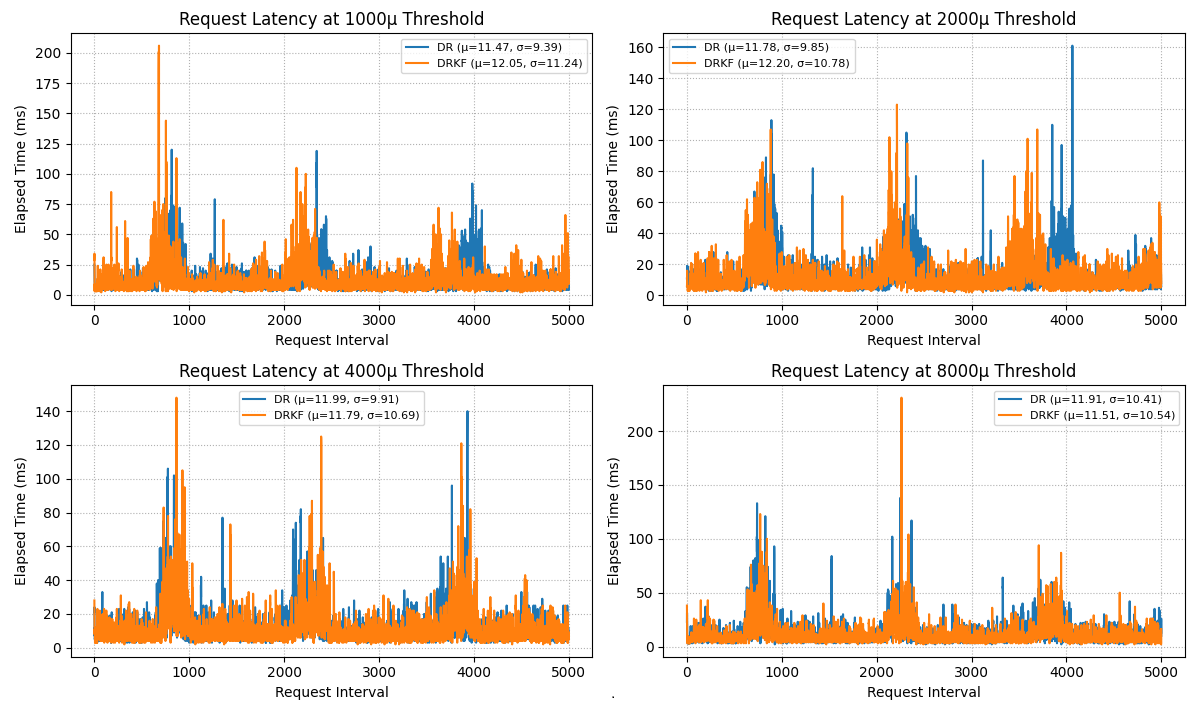}
\caption{$\mu$Bench Microservice Request Latency By Threshold on Compute Canada}
\label{fig:raw_elapsed_time}
\end{figure*}

In these experiments, the $\mu Bench$ serial 10-service application serves the stochastic workload for Drone Auto-scaler (DR) and Drone and Ksurf (DRKF) auto-scaler performance comparison  \cite{kubernetes}. The auto-scalers are configured with CPU thresholds starting at 1 millipod and increasing to an empirically selected maximum of 128 millipods as shown in Figure \ref{fig:mean_elapsed_time}. \\

\begin{figure}[H] 
\centering
\includegraphics[clip,width=1.0\linewidth]{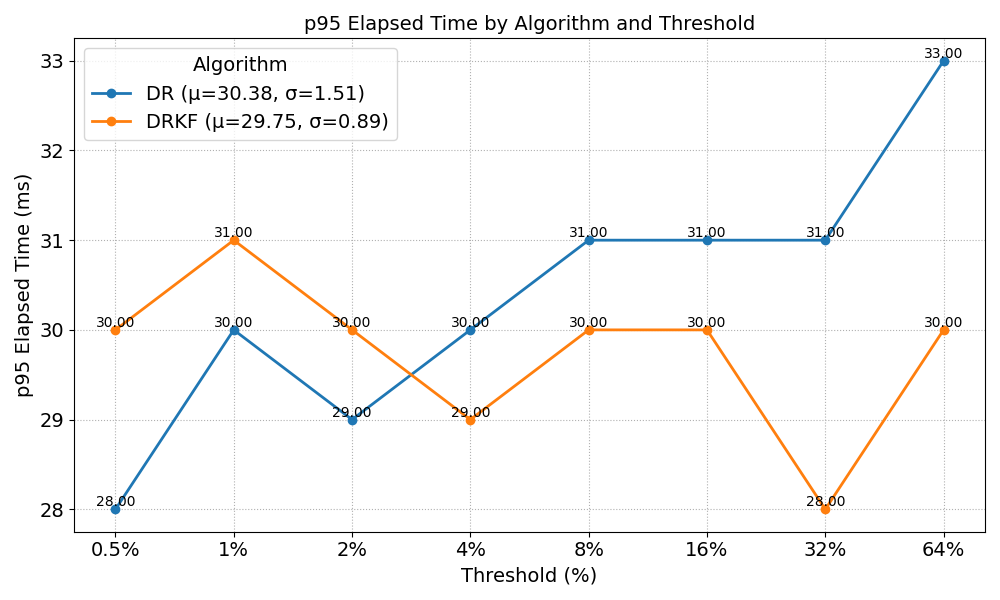}
\caption{Compute Canada $\mu$Bench p95 Request Latency By Threshold}
\label{fig:p95_elapsed_time}
\end{figure}

 Mean latency is shown to be 50 $\mu s$ lower in the DRKF auto-scaler over all thresholds decreasing from 480 $\mu s$ higher at 0.5 millipods to $530 \mu s$ lower at 64 millipods demonstrating the variance minimization of DRKF making it less reactive to jittery threshold crossing evident at low thresholds, while providing clean predictive data to Drone leading to higher performance as thresholds increase beyond $2\%$ as shown in Figure \ref{fig:raw_elapsed_time}. \\

\begin{figure}[H]
\centering
\includegraphics[clip,width=1.0\linewidth]{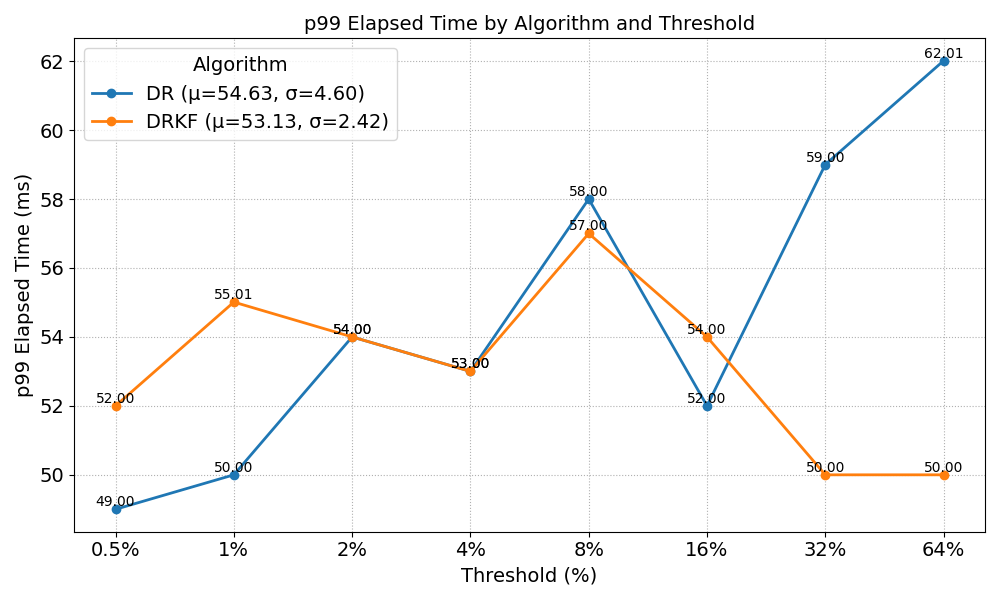}
\caption{Compute Canada $\mu$Bench p99 Request Latency By Threshold}
\label{fig:p99_elapsed_time}
\end{figure}

DRKF improves latency at moderate to high thresholds, including the request latency at p95 and p99 due to the Kalman filter having more time to converge leading to $3\%$ lower latency with $41\%$ lower latency variance at p95 and $2\%$ lower latency with $47\%$ lower latency variance at p99 as shown in Figure \ref{fig:p95_elapsed_time} and Figure \ref{fig:p99_elapsed_time}. We can model service latency as a Lipshitz function of the state $l_t=f(a_t,x_t)+\eta_t$ demonstrating the tail latency gain of Corollary \ref{Latency Corollary} \cite{QueuesLipshitzWhitt2018time}. \\

One future improvement to enhance performance of DRKF at low thresholds is to increase the process noise $Q$ and reduce the measurement noise $R$ for lower thresholds to make DRKF more sensitive to changes in measurements \cite{kfapplication}. \\

\begin{figure*}[h]
    \centering
    \begin{adjustbox}{width=0.85\linewidth} 
    \begin{tikzpicture}

        \node (master) [rectangle, draw, minimum width=5cm, minimum height=3cm, anchor=south] 
        at (0, 7) {};
        \node at ([yshift=0.2cm]master.south) {\textbf{Master Node}}; 
        
        \node (ingress) [below=0.5cm of master.north] {\includegraphics[width=1.5cm]{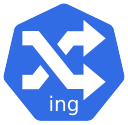}};

        \node (producer1) [rectangle, draw, minimum width=4cm, minimum height=2.5cm, left=4cm of ingress] {};
        \node at ([yshift=0.2cm]producer1.south) {\textbf{Producer 1}};
        \node (cronjob_p1) [above=0.5cm of producer1.south] {\includegraphics[width=1.2cm]{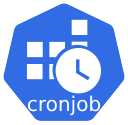}};

        \node (producer2) [rectangle, draw, minimum width=4cm, minimum height=2.5cm, right=4cm of ingress] {};
        \node at ([yshift=0.2cm]producer2.south) {\textbf{Producer 2}};
        \node (cronjob_p2) [above=0.5cm of producer2.south] {\includegraphics[width=1.2cm]{cronjob-128.png}};

        \node (worker1) [rectangle, draw, minimum width=7.2cm, minimum height=5cm, anchor=south, below left=3cm and 7cm of master] {};
        \node at ([yshift=0.2cm]worker1.south) {\textbf{Worker Node 1}};

        \node (worker2) [rectangle, draw, minimum width=7.2cm, minimum height=5cm, anchor=south, below right=3cm and 7cm of master] {};
        \node at ([yshift=0.2cm]worker2.south) {\textbf{Worker Node 2}};

        \node (worker3) [rectangle, draw, minimum width=7.2cm, minimum height=5cm, anchor=north, below=3cm of worker1] {};
        \node at ([yshift=0.2cm]worker3.south) {\textbf{Worker Node 3}};

        \node (worker4) [rectangle, draw, minimum width=7.2cm, minimum height=5cm, anchor=north, below=11cm of master] {};
        \node at ([yshift=0.2cm]worker4.south) {\textbf{Worker Node 4}};

        \node (worker5) [rectangle, draw, minimum width=7.2cm, minimum height=5cm, anchor=north, below=3cm of worker2] {};
        \node at ([yshift=0.2cm]worker5.south) {\textbf{Worker Node 5}};

        \foreach \i in {1,2} {
            \node (service\i) [below=0.5cm of master.south, xshift=-15cm + 10*\i cm] {\includegraphics[width=1.5cm]{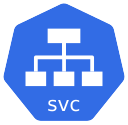}};
        }

        \foreach \i in {3,4,5} {
            \node (service\i) [below=5.5cm of master.south, xshift=-16cm + 4*\i cm] {\includegraphics[width=1.5cm]{svc-128.png}};
        }

        \foreach \i in {1,2} {
            \node (autoscaler\i) [rectangle, fill=green, draw, minimum width=3cm, minimum height=2cm, below=0.5cm of service\i] {};
            \node at ([yshift=-0.2cm]autoscaler\i.north) {\textbf{Drone / Autoscaler}};
            \node (hpa\i) [below=0.3cm of autoscaler\i.north, xshift=-0.8cm] {\includegraphics[width=1.2cm]{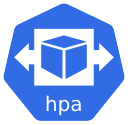}};
            \node (cronjob\i) [below=0.3cm of autoscaler\i.north, xshift=0.8cm] {\includegraphics[width=1.2cm]{cronjob-128.png}};
        }

        \foreach \i in {3,4,5} {
            \node (autoscaler\i) [rectangle, fill=green, draw, minimum width=3cm, minimum height=2cm, below=0.5cm of service\i] {};
            \node at ([yshift=-0.2cm]autoscaler\i.north) {\textbf{Drone / Autoscaler}};
            \node (hpa\i) [below=0.3cm of autoscaler\i.north, xshift=-0.8cm] {\includegraphics[width=1.2cm]{hpa-128.png}};
            \node (cronjob\i) [below=0.3cm of autoscaler\i.north, xshift=0.8cm] {\includegraphics[width=1.2cm]{cronjob-128.png}};
        }

        \foreach \i in {1,2,3,4,5} {
            \foreach \j in {1,2,3,4,5} {
                \node (pod\i\j) [below=1cm of worker\i.north, xshift=-3.5cm + 1.3*\j cm] {\includegraphics[width=1.2cm]{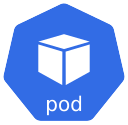}};
            }
        }

        \foreach \i in {1,2,3,4,5} {
            \foreach \j in {6,7,8,9,10} {
                \node (pod\i\j) [below=2.3cm of worker\i.north, xshift=-10.5cm + 1.3*\j cm] {\includegraphics[width=1.2cm]{pod-128.png}};
            }
        }

        \draw[->] (producer1) -- (ingress);
        \draw[->] (producer2) -- (ingress);

        \foreach \i in {1,2,3,4,5} {
            \draw[->] (ingress) -- (service\i);
            \draw[->] (service\i) -- (autoscaler\i);
            \draw[->] (autoscaler\i) -- (hpa\i);
            \draw[->] (autoscaler\i) -- (cronjob\i);
        }

        \foreach \i in {1,2,3,4,5} {
            \foreach \j in {1,2,3,4,5} {
                \draw[->] (autoscaler\i) -- (pod\i\j);
            }
        }

    \end{tikzpicture}
    \end{adjustbox}
    \caption{VarBench Kubernetes Cluster With 5 HPA \& Auto-scaler Nodes, 50 Worker Pods and 2 Producer Nodes}
    \label{fig:varbench_cluster}
\end{figure*}



\begin{figure*}[!htbp]
  \begin{adjustwidth}{-1cm}{-1cm}
\centering
\includegraphics[clip,width=0.95\linewidth]{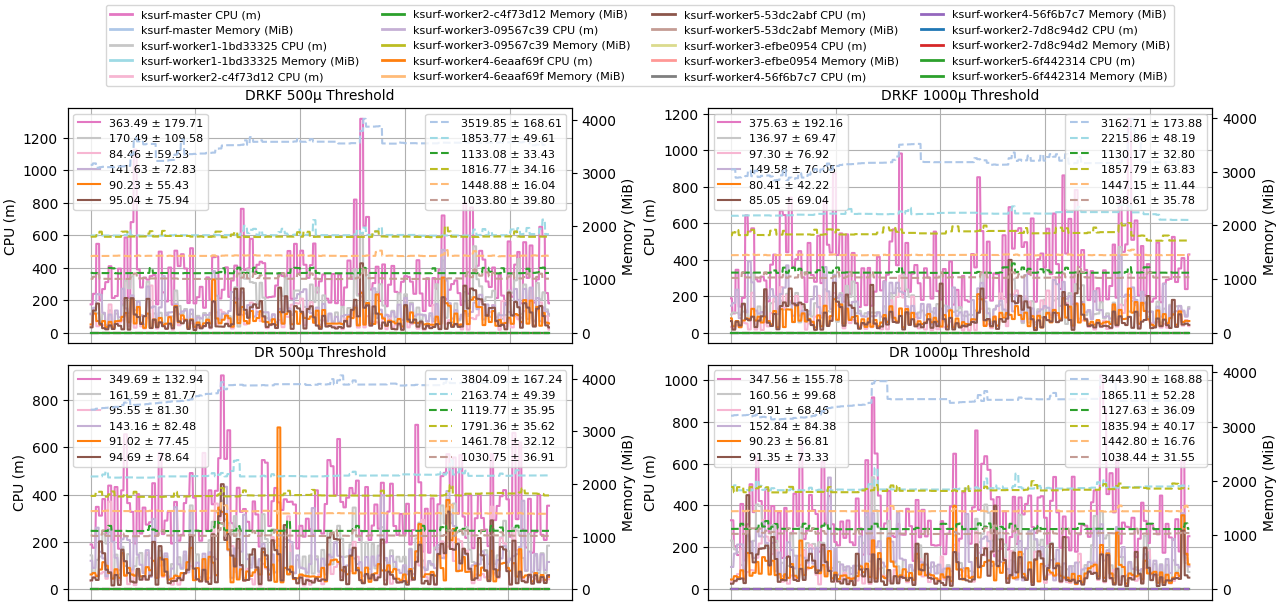}
\caption{Drone Node Metrics at 1 Millipod Threshold on Compute Canada}
\label{fig:drone_pod_metrics_1000mu}
  \end{adjustwidth}
\end{figure*}

\begin{figure*}[!htbp]
  \begin{adjustwidth}{-1cm}{-1cm}
\centering
\includegraphics[clip,width=0.95\linewidth]{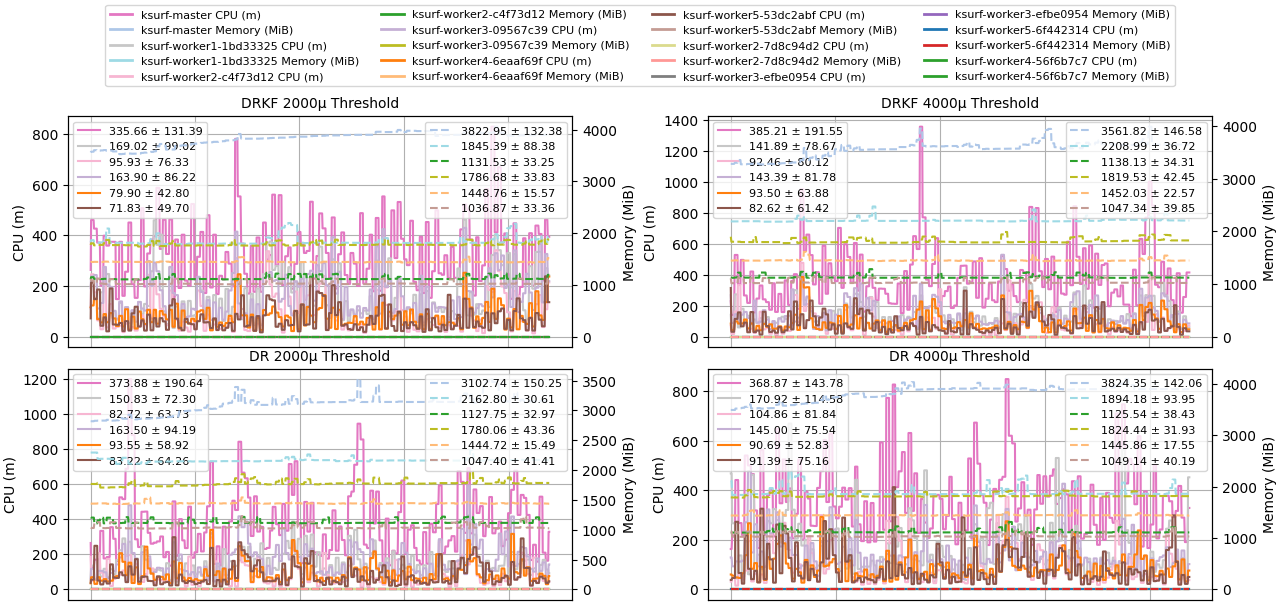}
\caption{Drone Node Metrics at 2 \& 4 Millipod Thresholds on Compute Canada}
\label{fig:drone_pod_metrics_4000mu}
  \end{adjustwidth}
\end{figure*}

\begin{figure*}[h]
\centering
\includegraphics[clip,width=0.9\linewidth]{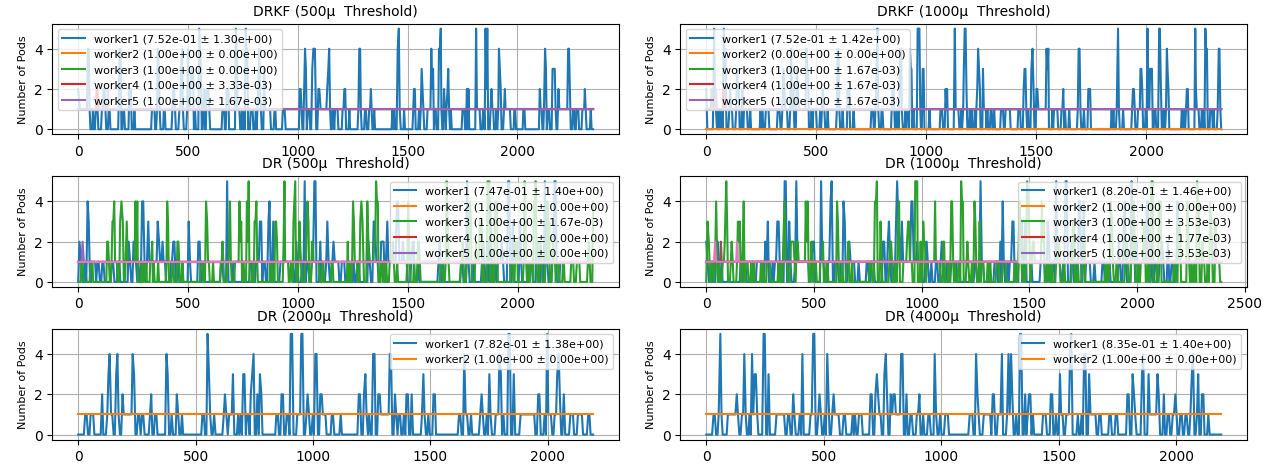}
\caption{Kafka Worker Pod Counts on Compute Canada}
\label{fig:varbench_pod_counts}
\end{figure*}

\begin{figure*}[h]
\centering
\includegraphics[clip,width=1.0\linewidth]{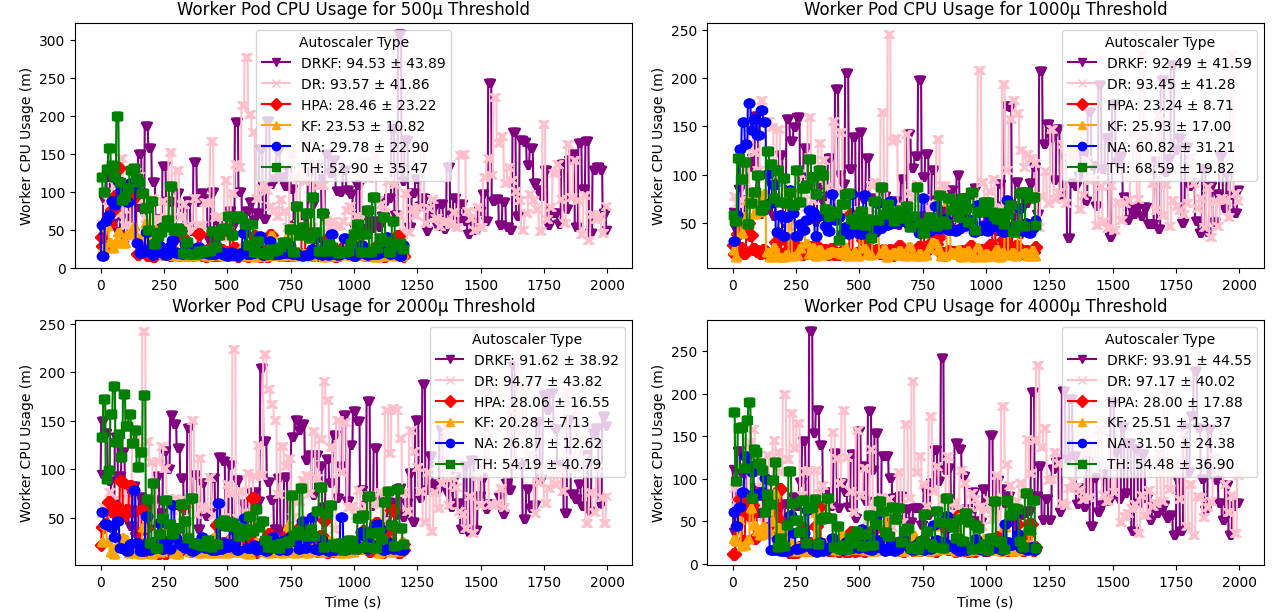}
\caption{Kafka Worker Pod CPU Usage By Threshold on Compute Canada \cite{Ksurf+}}
\label{fig:varbench_cpu_threshold}
\end{figure*}

\subsection{VarBench Kafka Benchmark Experiments}

$VarBench$ is a tool for benchmarking real-time message queue processing microservice applications with variable workloads at the application and underlying cloud platform layers \cite{KsurfConference}.  $VarBench$ consists of a Kubernetes cluster with master, scaler, producer and 50 worker pods, and an Apache Kafka application consisting of an Apache Zookeeper instance, 2 Kafka producer nodes, and consumer nodes hosted by Kubernetes worker pods with one consumer per pod, a Prometheus instance to collect metrics from worker pods, together with Drone for pod orchestration as shown in Figure~\ref{fig:varbench_cluster}  \cite{kubernetes} \cite{kafka}. \\

 \subsubsection{Kafka Workload \& Application}
 Typical online web traffic arrives in intervals that follow exponential distributions over small time scales fitting a Poisson process model \cite{PoissonWebAccess}. The experiments use a Poisson workload at $\lambda=0.125$, where the Kafka producer targets the Kafka service endpoint using a single topic with $n$ partitions and $n = num\_brokers$ brokers \cite{TwitterTrace} \cite{kfpcacode}. The Kafka client is a simple application running on the worker pods receiving the workload as notifications which are written to a log file \cite{kfpcacode}. \\

\subsubsection{$VarBench$ Configuration}
  The $VarBench$ configuration is augmented with a Prometheus pod instance metric collection and aggregation and a Drone instance for orchestration. The Drone instance is configured to use the private cloud algorithm with GP or Ksurf model depending on the experiment type \cite{zhang2023lifting}. The workload is driven by the producer nodes orchestrating all provisioned testing resources and compilation of test results. The testing goal is to compare DR to DRKF and four $VarBench$ auto-scaler types (HPA, KF, NA, TH) at selected CPU thresholds (0.5, 1, 2, 4 millipods) \cite{KsurfConference}. \\


\subsubsection{Results}
DRKF demonstrates a reduction of pod scaling actions regarding worker pods when compared to DR, leading to a $12.5\%$ reduction in worker pod count. DRKF displays lower worker pod variability represented by pod count variance reduction of over $110.1\%$ demonstrating the effectiveness of Ksurf at mitigating Drone cloud variability as shown in Figure~\ref{fig:varbench_pod_counts}. The overall performance improvement of DRKF over DR is shown in the worker node metrics where a significant reduction in CPU usage is shown with no significant difference in memory usage, demonstrated by a minimal $0.66\%$ increase in memory usage and a significant $7\%$ reduction in the CPU usage of the worker pod as shown in Figure \ref{fig:drone_pod_metrics_1000mu} and minimal $0.08\%$ increase in the memory usage of the worker pod and a significant reduction $4\%$ in the CPU usage of the worker pod in Figure \ref{fig:drone_pod_metrics_4000mu}. \\

\begin{figure*}[h]
\centering
\includegraphics[clip,width=1.0\linewidth]{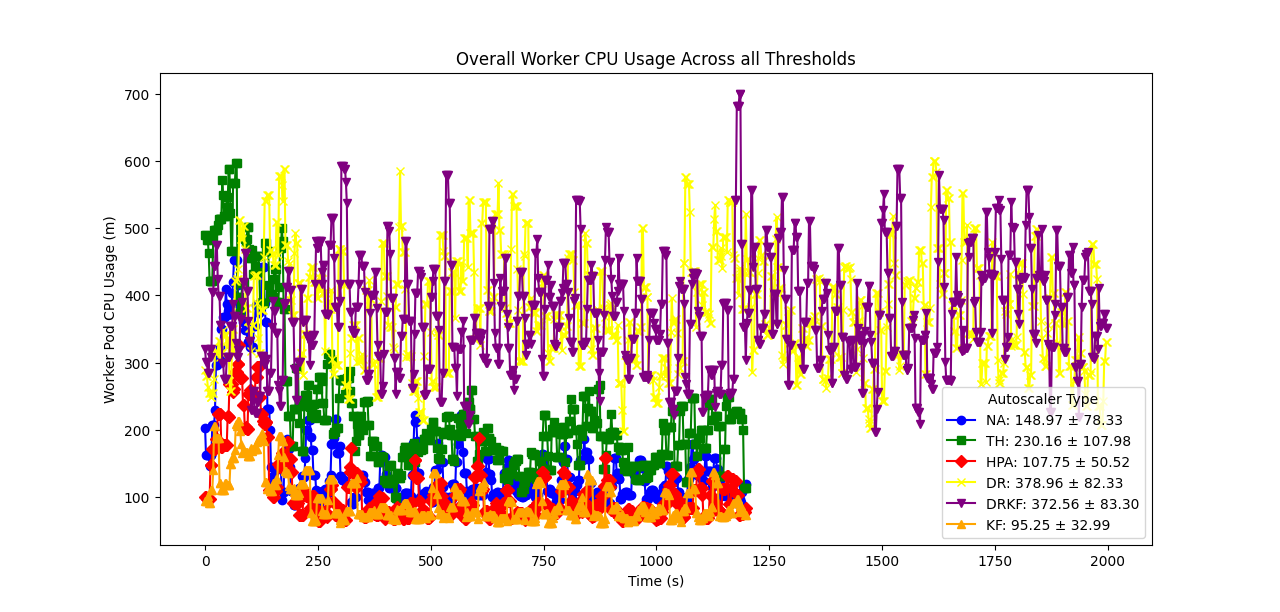}
\caption{Kafka Mean Worker Pod CPU Usage For All Thresholds on Compute Canada \cite{Ksurf+}}
\label{fig:varbench_cpu_all}
\end{figure*}

DRKF also demonstrates reduced resource variability demonstrated by a significant $3.81\%$ reduction in worker pod CPU usage and $7.74\%$ reduction in variance when compared to DR as shown in Figure \ref{fig:varbench_cpu_threshold}, Figure~\ref{fig:varbench_cpu_all} and Table~\ref{tbl:varbench_pod_summary}. The manually configured threshold-crossing $VarBench$ auto-scalers are more effective at maintaining lower worker pod CPU usage at all time periods demonstrated in Figure~\ref{fig:varbench_pod_counts}. Drone accurately models the performance-action relationship including broad sets of variables including past resource usage that is enhanced by lower variability predictions provided by Ksurf \cite{Ksurf+}. \\

DRKF mitigates a limitation of Drone regarding \textit{flash crowd} workloads which are characterized by short-term  bursty traffic which breaks the Gaussian Process prior assumption, due to the capability of Ksurf to support nonlinear models maintaining model consistency and reducing the risk of over-provisioning and under-provisioning under these conditions shown in Table \ref{tbl:varbench_pod_summary} and Table \ref{tbl:varbench_scaler_pod_summary}. \\

\begin{table}[h]
\centering
\begin{tabular}{ |p{2.0cm}|p{0.99cm}|p{0.99cm}|p{0.99cm}|p{0.99cm}|} 
 \hline
 \multicolumn{5}{|c|}{ CPU (Millipod) \& Memory (MiB) } \\
 \hline\hline
 Metric & DR Scaler CPU & DRKF Scaler CPU & DR Scaler Mem & DRKF Scaler Mem \\
 \hline
 Mean & 545.7 & 545.2 & 2788.5 & 2829.9 \\
 \hline
Std Deviation & 387.74 & 367.10 & 143.15 & 130.96 \\
 \hline
\end{tabular}
\caption{$VarBench$ Scaler Node Metric Summary}
\label{tbl:varbench_scaler_pod_summary}
 \end{table}

\begin{table}[h]
\centering
\begin{tabular}{ |p{2.0cm}|p{0.99cm}|p{0.99cm}|p{0.99cm}|p{0.99cm}|} 
 \hline
 \multicolumn{5}{|c|}{ CPU (Millipod) \& Memory (MiB) } \\
 \hline\hline
 Metric & DR Worker CPU & DRKF Worker CPU & DR Worker Mem & DRKF Worker Mem \\ 
 \hline
 Mean & 65.26  & \textbf{62.93} & 827.2 & 830.3 \\ 
 \hline
Std Deviation & 42.71 & \textbf{39.64} & 20.91 & 20.71 \\ 
 \hline
\end{tabular}
\caption{$VarBench$ Worker Pod Metric Summary}
\label{tbl:varbench_pod_summary}
 \end{table}

\subsection{Discussion}
\subsubsection{Worker Pod Memory and CPU Usage Tradeoff}
DRKF results in consistently higher memory usage on average across all worker pods in the variability experiments. However, the significance is 10x lower  than the DRKF CPU usage improvement demonstrating a minimal 0.373\% increase in memory usage and 0.96\% reduction in memory usage variance. In the private cloud scenario, Drone uses the CPU metric as the action variable to optimize memory as the reward variable \cite{zhang2023lifting}. Outliers in CPU usage data cause the Drone optimizer function to have fewer safe actions, leading to increased scaling, which impacts Drone GP more than DRKF due to the Ksurf attention filtering mechanism. The scaling of pods increases resource variability, the reduced scaling activity of DRKF is associated with reduced CPU and memory usage variance across all nodes as shown in Table ~\ref{tbl:varbench_scaler_pod_summary} and Table ~\ref{tbl:varbench_pod_summary}. For the Kafka application under test, reduced pod scaling activity leads to significantly lower CPU usage mean and variance due to worker pod application being more CPU than memory bottlenecked with CPU utilization 10x higher than memory utilization $63\% \gg 6\%$ shown in Table \ref{tbl:varbench_pod_summary}. \\

\subsubsection{Master and Scaler Node Resource Footprint}
DRKF has a 1.46\% higher scaler node memory footprint than DR because Ksurf is more memory intensive than GP on due to the transformer attention network memory footprint. The DRKF CPU footprint is lower due to Ksurf being more compute efficient leading 0.092\% lower CPU usage. Ksurf has  $O(n^3) \gg O(LN^2d)$ time complexity gain and $O(n^2) \gg O(N^2)$ memory complexity gain since GP batch size $n$ grows at each time step and the Ksurf data size $N$ is fixed $n \gg N$, however the memory used may be higher for Ksurf when the underlying implementation is operating in batch mode \cite{gregor2015draw} \cite{wenger2024computation}. This is a reasonable cost with regards to the significant 3.81\% CPU usage reduction in worker pods. The key lesson is that online methods including Ksurf in DRKF are better suited to real-time time series inference than batch learning algorithms as demonstrated by GP in DR when used as contextual bandit optimizer functions. \\

\section{Conclusions}

In this work, the Ksurf-Drone orchestrator is introduced to tackle the problem of container orchestration under conditions of high variability. Ksurf-Drone is compared to Drone and baseline auto-scaler algorithms on a set of experiments using the $VarBench$ benchmark. Ksurf uses a pre-trained attention layer and dimensionality reduction to mitigate outliers and noise in time series data \cite{pcabg} \cite{attentionNetworks}. A key benefit of Drone is its use of a non-parametric surrogate objective function GP, which makes it easier to implement and deploy, and provides robustness to user error introduced when selecting initial parameters. \\ 

The $\mu Bench$ benchmark experiments demonstrate the benefits of Ksurf as an optimization function for Drone using real-world application models with stochastic workloads. DRKF demonstrates $3\%$ lower p95 latency with $41\%$ lower p95 latency variance and $47\%$ lower p99 latency variance due to theoretical regret convergence guarantees illustrated in Lemma \ref{Regret lemma} and Corollary \ref{Latency Corollary}.  \\

The $VarBench$ benchmark experiments show the impact of the Ksurf predictor on the orchestration performance of Drone. Although the Kubernetes scaler node hosting Drone showed 1.48\% increased memory usage, the CPU usage was a slightly lower 0.09\% with 90\% reduction in CPU usage variance. The worker pods demonstrated a significant reduction in mean CPU usage of 3.59\% and corresponding reduction in CPU usage variance of 7.18\%. The reductions in memory and CPU usage variance reduce the probability over- and under-provisioning. This is captured in a 6.89\% reduction in average worker pod count for Ksurf-Drone. The reduction in worker pod CPU usage demonstrates significant positive impact of Ksurf on the orchestration quality of Drone by increasing robustness to variability in resource metrics used for orchestration decisions. The key lesson is that online estimation methods represented by Ksurf are more accurate than batch learning methods represented by GP on real-time short-horizon time series inference tasks. Also, the scaling period of the HPA, TH and KF auto-scalers defined by the worker CPU usage convergence duration after scaling actions is approximately half as long, indicating the effect of the higher computation time of each iteration of Drone. \\ 

Future work involves adapting Ksurf for graphical world models involving expanding KsurfNet using graph neural networks to complete model parameters of the state transition function and Kalman gain with a graph filter, enabling minimization of prediction error in estimating the hidden states of the microservice dependency graph structure from partial measurements \cite{GSPKalmanNet} \cite{KalmanNet2022}. Graphical KsurfNet would be instrumental in addressing this key limitation in contextual bandit and reinforcement learning approaches for microservice-oriented resource management \cite{zhang2023lifting}.


\newpage




\begin{thebibliography}{10}
\providecommand{\url}[1]{#1}
\csname url@samestyle\endcsname
\providecommand{\newblock}{\relax}
\providecommand{\bibinfo}[2]{#2}
\providecommand{\BIBentrySTDinterwordspacing}{\spaceskip=0pt\relax}
\providecommand{\BIBentryALTinterwordstretchfactor}{4}
\providecommand{\BIBentryALTinterwordspacing}{\spaceskip=\fontdimen2\font plus
\BIBentryALTinterwordstretchfactor\fontdimen3\font minus \fontdimen4\font\relax}
\providecommand{\BIBforeignlanguage}[2]{{%
\expandafter\ifx\csname l@#1\endcsname\relax
\typeout{** WARNING: IEEEtran.bst: No hyphenation pattern has been}%
\typeout{** loaded for the language `#1'. Using the pattern for}%
\typeout{** the default language instead.}%
\else
\language=\csname l@#1\endcsname
\fi
#2}}
\providecommand{\BIBdecl}{\relax}
\BIBdecl

\bibitem{zhang2023lifting}
Y.~Zhang, T.~Zhang, G.~Zhang, and H.-A. Jacobsen, ``Lifting the fog of uncertainties: Dynamic resource orchestration for the containerized cloud,'' in \emph{Proceedings Of The 2023 ACM Symposium On Cloud Computing}, 2023, pp. 48--64.

\bibitem{eismann2020microservices}
S.~Eismann, C.-P. Bezemer, W.~Shang, D.~Okanovi{\'c}, and A.~Van~Hoorn, ``Microservices: A performance tester's dream or nightmare?'' in \emph{Proceedings of the ACM/SPEC international conference on performance engineering}, 2020, pp. 138--149.

\bibitem{luo2021characterizingAlibaba}
S.~Luo, H.~Xu, C.~Lu, K.~Ye, G.~Xu, L.~Zhang, Y.~Ding, J.~He, and C.~Xu, ``Characterizing microservice dependency and performance: Alibaba trace analysis,'' in \emph{Proceedings of the ACM symposium on cloud computing}, 2021, pp. 412--426.

\bibitem{kabir2021uncertainty}
H.~D. Kabir, A.~Khosravi, S.~K. Mondal, M.~Rahman, S.~Nahavandi, and R.~Buyya, ``Uncertainty-aware decisions in cloud computing: Foundations and future directions,'' \emph{ACM Computing Surveys (CSUR)}, vol.~54, no.~4, pp. 1--30, 2021.

\bibitem{dean2013tail}
J.~Dean and L.~A. Barroso, ``The tail at scale,'' \emph{Communications of the ACM}, vol.~56, no.~2, pp. 74--80, 2013.

\bibitem{luo2022power}
S.~Luo, H.~Xu, K.~Ye, G.~Xu, L.~Zhang, G.~Yang, and C.~Xu, ``The power of prediction: microservice auto scaling via workload learning,'' in \emph{Proceedings of the 13th Symposium on Cloud Computing}, 2022, pp. 355--369.

\bibitem{awsSpotPricing}
\BIBentryALTinterwordspacing
``Amazon ec2 spot instances pricing– amazon web services.''\hskip 1em plus 0.5em minus 0.4em\relax AWS, 2023. [Online]. Available: \url{https://aws.amazon.com/ec2/spot/pricing/}
\BIBentrySTDinterwordspacing

\bibitem{awsBurstIntances}
\BIBentryALTinterwordspacing
``Burstable performance instances - Amazon Web Services.''\hskip 1em plus 0.5em minus 0.4em\relax AWS, 2023. [Online]. Available: \url{https://docs.aws.amazon.com/AWSEC2/latest/UserGuide/}
\BIBentrySTDinterwordspacing

\bibitem{shi2015clash}
J.~Shi, Y.~Qiu, U.~F. Minhas, L.~Jiao, C.~Wang, B.~Reinwald, and F.~{\"O}zcan, ``Clash of the titans: Mapreduce vs. spark for large scale data analytics,'' \emph{Proceedings of the VLDB Endowment}, vol.~8, no.~13, pp. 2110--2121, 2015.

\bibitem{ProScale2023}
C.~Ke, Z.~Sheng, T.~Chenghong, S.~Xiaohang, Y.~Zhaoheng, L.~Sanglu, L.~Yu, and G.~Qing, ``Proscale: Proactive autoscaling for microservice with time-varying workload at the edge,'' \emph{IEEE Trans. on Parallel and Distributed Systems}, vol.~34, no.~4, pp. 1294--1312, 2023.

\bibitem{autoscale2014}
T.~{Lorido-Botran}, J.~{Miguel-Alonso}, and J.~{Lozano}, ``A review of auto-scaling techniques for elastic applications in cloud environments,'' \emph{J Grid Computing}, vol.~12, p. 559–592, 2014.

\bibitem{DDoSFlashCrowd2011}
T.~Thapngam, S.~Yu, W.~Zhou, and G.~Beliakov, ``Discriminating ddos attack traffic from flash crowd through packet arrival patterns,'' in \emph{2011 IEEE Conference on Computer Communications Workshops (INFOCOM WKSHPS)}, 2011, pp. 952--957.

\bibitem{FlashCrowdEnsembleDetection2023althobaiti}
T.~Althobaiti, Y.~Sanjalawe, and N.~Ramzan, ``Securing cloud computing from flash crowd attack using ensemble intrusion detection system.'' \emph{Computer Systems Science \& Engineering}, vol.~47, no.~1, 2023.

\bibitem{yu2020microscaler}
G.~e.~a. Yu, ``Microscaler: Cost-effective scaling for microservice applications in the cloud with an online learning approach,'' \emph{IEEE Trans. on Cloud Computing}, vol.~10, no.~2, pp. 1100--1116, 2020.

\bibitem{ctxGBBandit}
\BIBentryALTinterwordspacing
H.~Zhang, J.~He, R.~Righter, Z.-J. Shen, and Z.~Zheng, ``Contextual gaussian process bandits with neural networks,'' in \emph{Advances in Neural Information Processing Systems}, A.~Oh, T.~Naumann, A.~Globerson, K.~Saenko, M.~Hardt, and S.~Levine, Eds., vol.~36.\hskip 1em plus 0.5em minus 0.4em\relax Curran Associates, Inc., 2023, pp. 26\,950--26\,965.

\bibitem{kfapplication}
Q.~{Li}, R.~{Li}, K.~{Ji}, and W.~{Dai}, ``Kalman filter and its application,'' in \emph{2015 8th International Conference on Intelligent Networks and Intelligent Systems (ICINIS)}, 2015, pp. 74--77.

\bibitem{kfEKFSpaceZhang}
J.~{Zhang}, ``Improvement and application of extended kalman filter algorithm in target tracking,'' \emph{Modern Computer}, vol.~32, pp. 11--16, 2012.

\bibitem{KsurfConference}
M.~Dang'ana and H.-A. Jacobsen, ``Ksurf: Attention kalman filter and principal component analysis for prediction under highly variable cloud workloads,'' in \emph{11th International Conference on Electrical Engineering, Computer Science and Informatics}, 2024, pp. 302--308.

\bibitem{alipourfard2017cherrypick}
O.~Alipourfard, H.~H. Liu, J.~Chen, S.~Venkataraman, M.~Yu, and M.~Zhang, ``$\{$CherryPick$\}$: Adaptively unearthing the best cloud configurations for big data analytics,'' in \emph{14th USENIX Symposium on Networked Systems Design and Implementation (NSDI 17)}, 2017, pp. 469--482.

\bibitem{banditsintroduction}
M.-A. Bandits, ``Introduction to multi-armed bandits.''

\bibitem{BanditsNoiseConfSetsLattimore}
T.~Lattimore and C.~Szepesvári, \emph{Bandit Algorithms}.\hskip 1em plus 0.5em minus 0.4em\relax Cambridge University Press, 2020.

\bibitem{rambo2021}
Q.~Li, B.~Li, P.~Mercati, R.~Illikkal, C.~Tai, M.~Kishinevsky, and C.~Kozyrakis, ``Rambo: Resource allocation for microservices using bayesian optimization,'' \emph{IEEE Computer Architecture Letters}, vol.~20, no.~1, pp. 46--49, 2021.

\bibitem{liu2019accordia}
Y.~Liu, H.~Xu, and W.~C. Lau, ``Accordia: Adaptive cloud configuration optimization for recurring data-intensive applications,'' in \emph{Proceedings of the ACM Symposium on Cloud Computing}, 2019, pp. 479--479.

\bibitem{MoRLMABGP}
D.~M. Roijers, L.~M. Zintgraf, P.~Libin, M.~Reymond, E.~Bargiacchi, and A.~Now{\'e}, ``Interactive multi-objective reinforcement learning in multi-armed bandits with gaussian process utility models,'' in \emph{Machine Learning and Knowledge Discovery in Databases}, F.~Hutter, K.~Kersting, J.~Lijffijt, and I.~Valera, Eds.\hskip 1em plus 0.5em minus 0.4em\relax Cham: Springer International Publishing, 2021, pp. 463--478.

\bibitem{KFGP}
J.~Hartikainen and S.~Särkkä, ``Kalman filtering and smoothing solutions to temporal gaussian process regression models,'' in \emph{2010 IEEE International Workshop on Machine Learning for Signal Processing}, 2010, pp. 379--384.

\bibitem{GP2014automatic}
D.~Duvenaud, ``Automatic model construction with gaussian processes,'' Ph.D. dissertation, University of Cambridge, 2014.

\bibitem{KalmanNet2022}
G.~Revach., S.~Nir, N.~Xiaoyong, E.~A. López, van Sloun. Ruud J.~G., and E.~Y. C., ``Kalmannet: Neural network aided kalman filtering for partially known dynamics,'' \emph{IEEE Trans. on Signal Processing}, vol.~70, pp. 1532--1547, 2022.

\bibitem{nnAGPB}
H.~Zhang, J.~He, R.~Righter, Z.-J. Shen, and Z.~Zheng, ``Contextual gaussian process bandits with neural networks,'' \emph{Advances in Neural Information Processing Systems}, vol.~36, pp. 26\,950--26\,965, 2023.

\bibitem{opolka2022adaptive}
F.~Opolka, Y.-C. Zhi, P.~Lio, and X.~Dong, ``Adaptive gaussian processes on graphs via spectral graph wavelets,'' in \emph{International Conference on Artificial Intelligence and Statistics}.\hskip 1em plus 0.5em minus 0.4em\relax PMLR, 2022, pp. 4818--4834.

\bibitem{GSPKalmanNet}
I.~Buchnik, G.~Sagi, N.~Leinwand, Y.~Loya, N.~Shlezinger, and T.~Routtenberg, ``Gsp-kalmannet: Tracking graph signals via neural-aided kalman filtering,'' \emph{IEEE Transactions on Signal Processing}, vol.~72, pp. 3700--3716, 2024.

\bibitem{prometheus}
\BIBentryALTinterwordspacing
``Prometheus.''\hskip 1em plus 0.5em minus 0.4em\relax SoundCloud, 2025. [Online]. Available: \url{https://prometheus.io}
\BIBentrySTDinterwordspacing

\bibitem{k8sResources}
\BIBentryALTinterwordspacing
``Resource management for pods and containers | kubernetes.''\hskip 1em plus 0.5em minus 0.4em\relax Kubernetes, 2025. [Online]. Available: \url{https://kubernetes.io/docs/concepts/configuration/manage-resources-containers/}
\BIBentrySTDinterwordspacing

\bibitem{LstmArimaHyperParameterGrid}
U.~Singh, S.~Tamrakar, K.~Saurabh, R.~Vyas, and O.~Vyas, ``Hyperparameter tuning for lstm and arima time series model: A comparative study,'' in \emph{2023 IEEE 4th Annual Flagship India Council International Subsections Conference (INDISCON)}, 2023, pp. 1--6.

\bibitem{KFComplexity}
M.~Raitoharju and R.~Piché, ``On computational complexity reduction methods for kalman filter extensions,'' \emph{IEEE Aerospace and Electronic Systems Magazine}, vol.~34, no.~10, pp. 2--19, 2019.

\bibitem{BanditsCtxLinearChu11a}
W.~Chu, L.~Li, L.~Reyzin, and R.~Schapire, ``Contextual bandits with linear payoff functions,'' in \emph{Proceedings of the Fourteenth International Conference on Artificial Intelligence and Statistics}, ser. Proceedings of Machine Learning Research, G.~Gordon, D.~Dunson, and M.~Dudík, Eds., vol.~15.\hskip 1em plus 0.5em minus 0.4em\relax Fort Lauderdale, FL, USA: PMLR, 11--13 Apr 2011, pp. 208--214.

\bibitem{BanditsEpsilonShrink2011abbasi}
Y.~Abbasi-Yadkori, D.~P{\'a}l, and C.~Szepesv{\'a}ri, ``Improved algorithms for linear stochastic bandits,'' \emph{Advances in neural information processing systems}, vol.~24, 2011.

\bibitem{BanditRegretVariance2024jia2024}
Z.~Jia, J.~Qian, A.~Rakhlin, and C.-Y. Wei, ``How does variance shape the regret in contextual bandits?'' \emph{Advances in Neural Information Processing Systems}, vol.~37, pp. 83\,730--83\,785, 2024.

\bibitem{krener2002convergence}
A.~J. Krener, ``The convergence of the extended kalman filter,'' in \emph{Directions in mathematical systems theory and optimization}.\hskip 1em plus 0.5em minus 0.4em\relax Springer, 2002, pp. 173--182.

\bibitem{QueuesLipshitzWhitt2018time}
W.~Whitt, ``Time-varying queues,'' \emph{Queueing models and service management}, vol.~1, no.~2, 2018.

\bibitem{gregor2015draw}
K.~Gregor, I.~Danihelka, A.~Graves, D.~Rezende, and D.~Wierstra, ``Draw: A recurrent neural network for image generation,'' in \emph{International conference on machine learning}.\hskip 1em plus 0.5em minus 0.4em\relax PMLR, 2015, pp. 1462--1471.

\bibitem{wenger2024computation}
J.~Wenger, K.~Wu, P.~Hennig, J.~Gardner, G.~Pleiss, and J.~P. Cunningham, ``Computation-aware gaussian processes: Model selection and linear-time inference,'' \emph{Advances in Neural Information Processing Systems}, vol.~37, pp. 31\,316--31\,349, 2024.

\bibitem{googleBenchmark}
D.~Bruening, Q.~Zhao, and R.~Kleckner, ``Google workload trace benchmark - charlie,'' \url{https://console.cloud.google.com/storage/browser/external-traces/charlie}, 2023, accessed: 2023-05-24.

\bibitem{computeCanada}
\BIBentryALTinterwordspacing
``Compute canada.''\hskip 1em plus 0.5em minus 0.4em\relax Digital Alliance Canada, 2025. [Online]. Available: \url{https://www.alliancecan.ca/en}
\BIBentrySTDinterwordspacing

\bibitem{GPKalmanRTS}
J.~Hartikainen and S.~Särkkä, ``Kalman filtering and smoothing solutions to temporal gaussian process regression models,'' in \emph{2010 IEEE International Workshop on Machine Learning for Signal Processing}, 2010, pp. 379--384.

\bibitem{MAMLKalmannet2025}
S.~Chen, Y.~Zheng, D.~Lin, P.~Cai, Y.~Xiao, and S.~Wang, ``Maml-kalmannet: A neural network-assisted kalman filter based on model-agnostic meta-learning,'' \emph{IEEE Transactions on Signal Processing}, vol.~73, pp. 988--1003, 2025.

\bibitem{muBench}
D.~Andrea, F.~Ludovico, and P.~Luca, ``$\mu Bench$: An open-source factory of benchmark microservice applications,'' \emph{IEEE Trans. on Parallel and Distributed Systems}, vol.~34, no.~3, pp. 968--980, 2023.

\bibitem{GloudAwsVariabilty}
A.~Iosup, N.~Yigitbasi, and D.~Epema, ``On the performance variability of production cloud services,'' in \emph{2011 11th IEEE/ACM International Symposium on Cluster, Cloud and Grid Computing}, 2011, pp. 104--113.

\bibitem{muBenchDeathStar}
M.~S. Rashmi., R.~Vismaya, P.~Abhishek, R.~G. Ranjitha, B.~Tushar, and A.~Prafullata, ``Synthetic to real-world: Insights on microservices-based application benchmarks,'' in \emph{2024 IEEE International Conference on Information Technology, Electronics and Intelligent Communication Systems (ICITEICS)}, 2024, pp. 1--8.

\bibitem{kubernetes}
``Kubernetes: Production-grade container orchestration,'' \url{https://kubernetes.io/}, 2024, version 1.29, Accessed: 2024-01-24.

\bibitem{Ksurf+}
M.~Dang’ana and H.-A. Jacobsen, ``Ksurf+: Attention kalman filter for prediction under highly variable cloud workloads,'' 2025.

\bibitem{kafka}
``{Apache Kafka},'' \url{http://kafka.apache.org/}, 2024, version 1.29, Accessed: 2024-01-24.

\bibitem{PoissonWebAccess}
\BIBentryALTinterwordspacing
H.~Shigeki, F.~Yoshiharu, S.~Masaya, T.~Masahiko, and Y.~Naoki, ``Web server access trend analysis based on the poisson distribution,'' in \emph{Proceedings of the 6th International Conference on Software and Computer Applications}, 2017, p. 256–261. [Online]. Available: \url{10.1145/3056662.3056701}
\BIBentrySTDinterwordspacing

\bibitem{TwitterTrace}
``{Twitter Developer API hashtag \#Ukraine},'' https://developer.twitter.com, accessed: 2023-05-24.

\bibitem{kfpcacode}
\BIBentryALTinterwordspacing
M.~{Dang'ana}, ``Akf-pca source code,'' \emph{Github}, 2025. [Online]. Available: \url{https://github.com/msrg/kfpca}
\BIBentrySTDinterwordspacing

\bibitem{pcabg}
S.~Wold, K.~Esbensen, and P.~Geladi, ``Principal component analysis,'' \emph{Chemometrics and Intelligent Laboratory Systems}, vol.~2, pp. 37--52, 1987.

\bibitem{attentionNetworks}
\BIBentryALTinterwordspacing
Y.~Kim, C.~Denton, L.~Hoang, and A.~M. Rush, ``Structured attention networks,'' \emph{International Conference on Learning Representations}, 2017. [Online]. Available: \url{https://console.cloud.google.com/storage/browser/external-traces/charlie}
\BIBentrySTDinterwordspacing

\end{thebibliography}

\end{document}